\documentclass[journal]{IEEEtran}
\usepackage{cite}
\usepackage{graphicx,amssymb,lineno}
\usepackage{epstopdf}
\usepackage{amsmath,amsfonts,amssymb}

\usepackage{algorithm}
\usepackage{algorithmic}
\usepackage[usenames]{color}
\usepackage{float}
\usepackage{mathtools}
\usepackage{cite}
\usepackage{booktabs}
\usepackage{stfloats}
\usepackage{multirow}
\usepackage{lipsum}
\usepackage{cuted}
\usepackage{bm}

\renewcommand{\algorithmicrequire}{\textbf{Initialization:}}

\usepackage{graphicx,graphics,color,epsfig,subfigure,graphpap,rotate}
\usepackage{times, verbatim, subfigure, epsfig, graphicx, latexsym, amsmath}
\usepackage{url}
\usepackage{subfigure}

\begin{document}

%\title{\huge{Resource Allocation for  Millimeter-wave Train-ground Communications in High-speed Railway Scenarios}}
\title{Deep Reinforcement Learning Coordinated\\ Receiver Beamforming for Millimeter-Wave Train-ground Communications}

\author{Xutao~Zhou,
        Xiangfei~Zhang,
        Chen~Chen,
        Yong~Niu,~\IEEEmembership{Member,~IEEE},
        Zhu Han,~\IEEEmembership{Fellow,~IEEE},
        He~Wang,
        Chengjun~Sun,
        Bo~Ai,~\IEEEmembership{Senior Member,~IEEE}
        and Ning~Wang,~\IEEEmembership{Member,~IEEE}
}

\maketitle

\begin{abstract}
As more and more people choose high-speed rail (HSR) as a means of transportation for short trips, there is ever growing demand of high quality of multimedia services. With its rich spectrum resources, millimeter wave (mm-wave) communications can satisfy the high network capacity requirements for HSR. Also, it is possible for receivers (RXs) to be equipped with antenna arrays in mm-wave communication systems due to its short wavelength. However, as HSRs run with high speed, the received signal power (RSP) varies rapidly over a cell and it is the lowest at the edge of the cell compared to other locations. Consequently, it is necessary to conduct research on RX beamforming for HSR in mm-wave band to improve the quality of the received signal. In this paper, we focus on RX beamforming for a mm-wave train-ground communication system. To improve the RSP, we propose an effective RX beamforming scheme based on deep reinforcement learning (DRL), and develop a deep Q-network (DQN) algorithm to train and determine the optimal RX beam direction with the purpose of maximizing average RSP.
Through extensive simulations, we demonstrate that the proposed scheme has better performance than the four baseline schemes in terms of average RSP at most positions on the railway.

\end{abstract}


%---------------------------
\section{Introduction}\label{S1}
%---------------------------

High-speed railway (HSR), with its high efficiency, convenience, safety, comfort and environmental protection and other characteristics, has become a new trend of future transportation development. In recent years, with the rapid development of HSR infrastructure construction, it has gradually become an important means of transportation in various countries. The length of China's railways in operation has reached nearly 39,000 km~\cite{r1-1}. On the other hand, since many passengers are accustomed to broadband wireless access in their daily life, more and more people hope to have high-quality broadband wireless access on mobile terminals.

However, in the HSR scenarios, the high-speed of train causes frequent handovers. When the cell radius is 1 to 2km, the train running at 350km/h will handover every 10 to 20s~\cite{r1-1}. In addition, the rapid relative movement between the train and ground base station (BS) causes more serious Doppler shift and smaller channel coherence time. So the wireless channel in HSR scenarios has obvious non-stationary and fast time-varying characteristics, which seriously reduces the performance of train-ground communication systems~\cite{r1-2}. Moreover, due to the complexity and non-stationarity of HSR scenarios, there are weak field strength areas and blind areas, and the train body of metal material causes great penetration loss to the signal from the BS~\cite{r1-3}.
%Therefore, meeting passengers' compelling demand for broadband mobile communications in the HSR environment has become a key technical challenge. It has become particularly important to carry out research on broadband wireless communication technologies in HSR scenarios.
At the same time, with the emergence of various entertainment applications, the demand rate may easily reach 0.5-5Gbps in future HSR communication system~\cite{fiber}. As a result, the current wireless transmission scheme will not be able to meet the demands of HSR passengers.

For this purpose, the 30-300GHz millimeter-wave (mm-wave) frequency band has been favored by many researchers.
Owing to its wide bandwidth and rich spectrum resources, it is not only consistent with the current development trend of wireless communication systems, but also can satisfy multi-gigabit wireless services such as online gaming, video calling and so on~\cite{J1}, which the current communication systems for railways like Global System for Mobile Communications for Railways (GSM-R) and LTE-R can't provide. Thus, mm-wave communications are regarded as a promising technique to meet the growing network capacity requirements caused by the rapid development in the HSR industry~\cite{channel}\cite{J8}.
To date, in some countries, mm-wave have been applied in HSR communication systems, like the Japanese Shinkansen, the TVE test line of maglev train in Germany and the maglev train in Shanghai, China~\cite{2018Application}.

Unfortunately, there is severe penetration loss to received signal power (RSP) due to the train's body is made of metal material. To avoid this problem, mobile relays (MRs) are introduced. There is a two-hop communication link between the BS and users on the train, which can effectively avoid the adverse effects of penetration loss and group handovers~\cite{2019A}.
And it is important to note that the communication link is formed between the BS and MR in this paper.
Besides, although there are many benefits to take mm-wave communication into practice, a challenge that cannot be ignored is its high carrier frequency, which causes higher path loss than lower frequency bands. Take 60 GHz frequency band for example, there is 28 decibels (dB) higher free space path loss than that at 2.4GHz~\cite{2011Interference}.
To address this problem, directional antennas are introduced to compensate the severe link attenuation of mm-wave communications and utilize the beamforming technology for high antenna gain~\cite{2014Millimeter}.
In previous HSR communication systems that operate at microwave bands, the transmitter (TX) (i.e., the trackside BS) is equipped with antenna arrays to form space-orthogonal beams while the RX usually has a single antenna since the space is limited. Therefore, there is a problem that the receiver (RX) (i.e., MR) has to frequently feed back its estimated downlink channel state information (CSI) to BSs and wait for the TX beam adjustments in order to enhance the quality of received signal.
The link adaption approach that based on frequent feedback, however, will not be suitable for fast-varying channels under high mobility scenarios.

Unlike the conventional space division multiple access (SDMA) systems where only TX has sufficient space to equip with large antenna arrays, thanks to the short wavelength of mm-wave, the RX can also accommodate large antenna arrays to form beams. Consequently, the RX side can autonomously adjust beams to adapt to fast-varying wireless channels, thereby avoiding the problems mentioned above.
Also, as one of the most typical characteristics of HSR is its high speed, which causes frequent handovers and severe Doppler shift, the RSP varies rapidly over a cell and it is the lowest at the edge of the cell compared to other locations.
Consequently, it is significant to conduct research on RX beamforming to improve the quality of received signal.
Moreover, beam training overhead is also a significant factor influencing the system performance. Exhaustive beam search (EBS) is a conventional scheme that traverses all possible beam directions based on the codebook design, which will cause high latency and severe misalignment especially in the HSR scenario. Therefore, a suboptimal but more practical scheme with low overhead and less demand of CSI and other information should be proposed in significant needs.

With the development of computer science, machine learning has become a very popular technology today. As an important branch in the field of machine learning, reinforcement learning (RL) is independent of supervised learning and unsupervised learning. RL algorithm can realize intelligent decision-making by constantly interacting with the environment and using environmental feedback for self-optimization and self-learning.
In addition, deep reinforcement learning (DRL) that combines deep learning (DL) and RL has been proposed recently~\cite{2015Human}. This promising technique is used to solve the situation where there are enormous state-action pairs, for which it is hard to find the optimal policy between the agent and the environment within a finite number of interaction steps. And it is widely used in the decision-making process of solving decision-making problems with large state-action spaces~\cite{2020Future}~\cite{2019Model}. The great advantage of DRL is its ability to make decisions quickly after training.

In this paper, we conduct a research on the RX beamforming mechanism for the mm-wave train-ground communication system where the MRs are also equipped with large antenna arrays. We concentrate on enhancing the RSP by employing DRL to RX beamforming.
We investigate an effective DRL based RX beamforming method for mm-wave train-ground communications, which allows MRs to choose the
most suitable beam direction from its local information of the fast-varying environment in time.
Based on the DRL algorithm, our scheme achieves lower overhead makes the MR determine the beam direction more quickly than the EBS scheme. As a consequence, integrating mm-wave communication and DRL into HSR communication scenarios and the RX beamforming scheme stated in this paper will be of great value for improving the RSP.
The main contributions of this paper are summarized as follows.

\begin{itemize}
\item We focus on the RX beamforming problem of mmWave train-ground communication system where the RX is also equipped with large antenna arrays, aiming at optimizing the RX beamforming problem in terms of average RSP. Then we formulate the optimization problem as a mixed integer nonlinear programming problem.

\item By taking advantages of DRL, we propose an efficient DQN based algorithm for RX beamforming to achieve better system performance through meticulously designing state, action and reward. Next we formulate the objective of the proposed scheme as maximizing average RSP. In our proposed scheme, in coverage of a TX beam, the RX side can autonomously adjust RX beam directions rather than frequently feeding back CSI to TX, which greatly reduces overheads.

\item By conducting simulations under location of the train, we demonstrate that the proposed DQN based RX beamforming scheme
    outperforms the two baseline schemes, i.e., FBA and 16-beams DQN algorithms at every positions on the railway and most positions comparing with $\gamma$-greedy algorithm.
\end{itemize}

The rest of the paper is organized as follows. In Section~\ref{S2}, we
provide
%give a detailed
an detailed overview of the related work. In Section~\ref{S3}, we introduce the
%established
mm-wave train-ground communication system model, and formulate the problem of of maximizing average RSP of MR.
%The construction of Lagrange function and the basic ideas of the introduced SQP algorithm are given
The proposed DQN algorithm is presented
in Section~\ref{S4} and evaluated in Section~\ref{S5}. Finally, Section~\ref{S6} concludes this paper.
%the conclusions are drawn in Section~\ref{S6}.

%---------------------------
\section{Related Work}\label{S2}
%---------------------------

There has been quite a few research on mm-wave communications for HSRs.
In~\cite{J20}~\cite{J21}, channel measurement was conducted at mm-wave frequency band in different HSR scenarios. Channel characteristics, such as path loss, K-factor, Doppler shifts, coherence time and so on, were investigated under the help of a 3-D ray tracer (RT).
Also, in~\cite{2018Clustering,2020Trajectory,2019Train}, some original channel modeling methods were introduced to describe the time varying wireless channel more precisely and prove the feasibility of mm-wave communications. In~\cite{J8}, several trackside mm-wave network architectures where deployed both orthogonal frequency-division multiplexing (OFDM) and single carrier (SC) technology were designed in order to overcome challenges in HSR scenarios.
In addition, some higher mm-wave frequency bands, such as the 30GHz~\cite{2019Train} and the 24-40GHz band~\cite{2017Mobile} were studied to increase the transmission capacity to a greater extent in train-ground communications.

There is also considerable work on mm-wave beamforming for HSRs.
Channel simulation was conducted at 60GHz frequency band in order to design a patch antenna array and beamforming scheme for HSRs.
Xu \emph{et al.} studied a joint TX-RX beamforming and power allocation problem to maximize the system sum rate for the train-ground communications~\cite{BF}.
Yu \emph{et al.}~\cite{FBF1} and Kim \emph{et al.}~\cite{FBF2} investigated the gap of system performance between fixed beamforming and adaptive beamforming  in terms of the received signal strength, root-mean-squared delay spread, and so on. Results showed that the gap is smaller in the HSR tunnel than in the train station owing to the severe beam misalignment.

To keep the received signal to noise ratio (SNR) at a relatively high and stable value and reduce feedback overheads, authors in~\cite{J7} proposed a scheme with both long-term TX beamforming and short-term RX beamforming.
Given the stability and reliability of the mm-wave HSR communication system, in~\cite{LF} and~\cite{Steady}, some novel beamforming schemes under an interleaved redundant coverage architecture were provided.
In~\cite{J7}, regarding the problem of train safety, the authors introduced an mm-wave beamforming method for disaster radar detection in HSR scenarios.
In addition, network capacity is a key index to measure the performance of the communication system. In~\cite{adaptive}, the authors presented an adaptive mm-wave multi-beamforming scheme where multiple beams with different beamwidth were used by the BS at the same time. The simulation results showed that an optimal system capacity can be maintained.
Since mm-wave communication has the characteristic of severe propagation attenuation and vulnerability blockage, an efficient hybrid beamforming approach was introduced to combat the blockage problem~\cite{2020Efficient}.

Recently, DRL has caused much attention both in academic and industry. As a result, a lot of research that applying DRL to wireless communications has been conducted.
Luong \emph{et al.}~\cite{survey} presented a comprehensive survey on applications of DRL in communications and networking, such as the Internet of Things (IoT) and unmanned aerial vehicle (UAV) networks.
With the development of intelligent reflecting surface (IRS) technology, a novel DRL-based scheme, which jointly optimized BS's beamforming and IRS's reflecting beamforming was put forward to enhance security of the communication system~\cite{IRS}.
Taking energy efficiency (EE) into consideration, the authors in~\cite{cell} proposed an uplink beamforming scheme based on DRL under the architecture of cell-free network to achieve maximum EE.
In~\cite{Ge}, the authors proposed a DRL-based distributed downlink beamforming method, where each BS can train its own deep Q-network (DQN) to determine the most suitable beam direction with limited information. This scheme can effectively reduce system overhead and achieved near-optimal system performance.

In terms of the high mobility scenario, DRL has also been adopted for optimizing congestion control performance and radio resource management.
In~\cite{CC}, due to the poor performance of traditional Transmission Control Protocol (TCP) in the HSR scenario, the authors introduced a DRL-based congestion control algorithm, where a DQN was applied to precisely control congestion window.
Xu \emph{et al.}~\cite{PA} proposed an algorithm based on DRL, which makes the agent, i.e., the MR learn the decisions of power allocation from the past experience with the aim of maximizing the system throughput.
As for high mobility heterogeneous network, authors in~\cite{RA} investigated the resource allocation problem and presented a DRL-based Time Division Duplex (TDD) configuration algorithm to decrease packet loss rate.

To sum up, all these works mentioned above do not combine mm-wave communication, RX beamforming and DRL together in the HSR scenario.
Therefore there is considerable space for performance improvement in terms of RSP.
In this paper, we study the RX beamforming problem in the mm-wave train-ground communication system, and propose an algorithm based on DRL to achieve excellent performance on RSP.

%---------------------------
\section{System Overview and Problem Formulation}\label{S3}
%---------------------------

%---------------------------
\subsection{System Model}\label{S3-1}
%---------------------------

\begin{figure}[t]
    \centering
    \subfigure[Network operating in the unidirectional mode.] {\includegraphics*[width=1\columnwidth,height=0.7in]{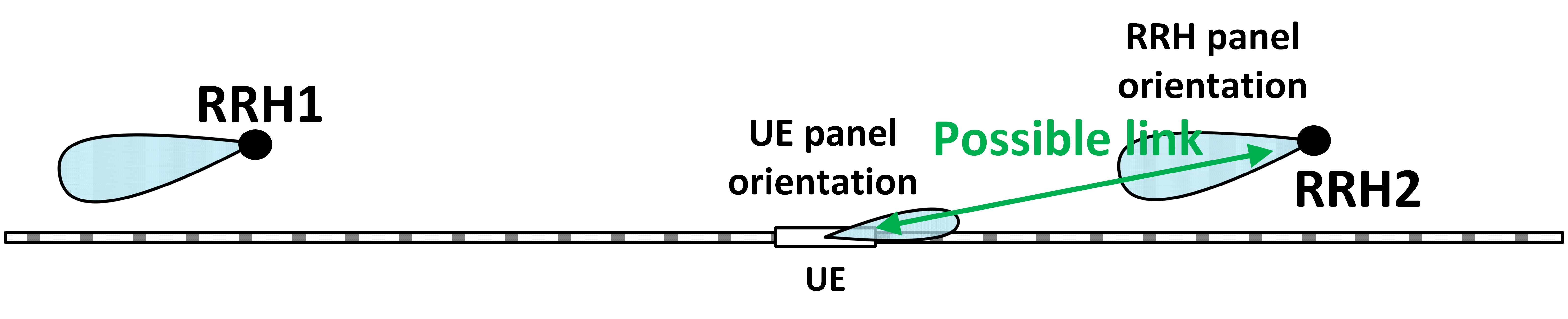}}
    \subfigure[Network operating in the bidirectional mode.] {\includegraphics*[width=1\columnwidth,height=0.7in]{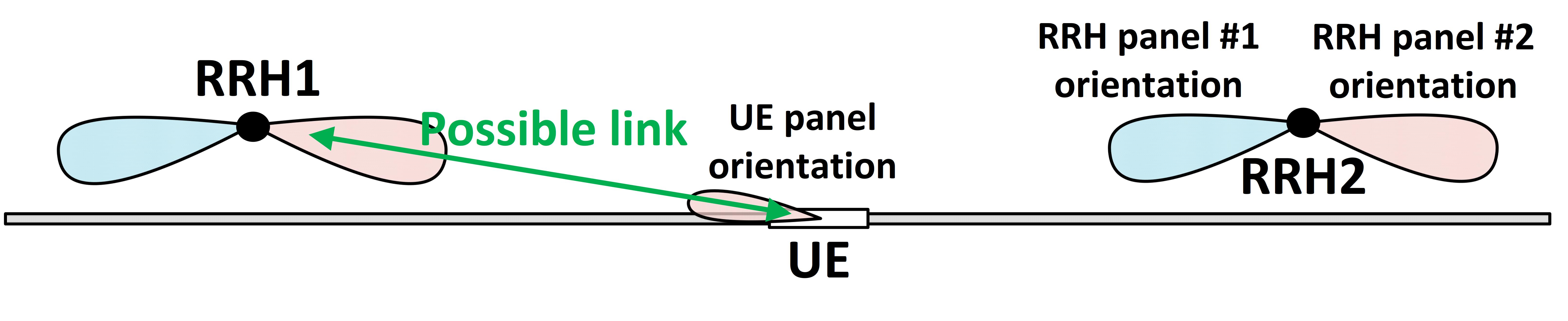}}
    \subfigure[Both network and UE support bidirectional operation.] {\includegraphics*[width=1\columnwidth,height=0.7in]{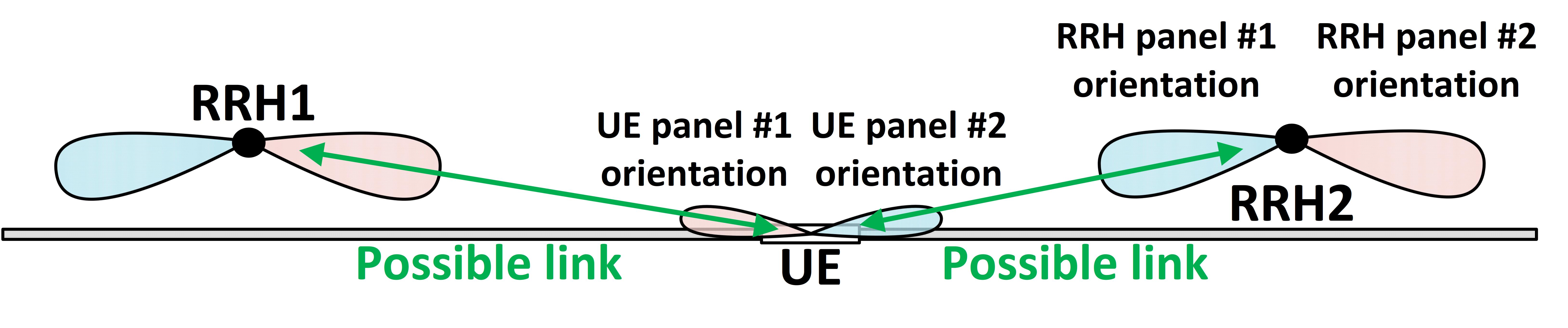}}
    \caption{Three scenarios of SFN deployment.}
    \label{fig1}
\end{figure}

%\begin{figure}[t]
%\begin{center}
%\includegraphics*[width=1\columnwidth,height=0.7in]{fig1.jpg}
%\end{center}
%\caption{Network operating in unidirectional mode.} \label{fig1}
%\end{figure}
%\begin{figure}[t]
%\begin{center}
%\includegraphics*[width=1\columnwidth,height=0.7in]{fig2.jpg}
%\end{center}
%\caption{Network operating in bidirectional mode.} \label{fig2}
%\end{figure}
%\begin{figure}[t]
%\begin{center}
%\includegraphics*[width=1\columnwidth,height=0.7in]{fig3.jpg}
%\end{center}
%\caption{Both network and UE support bidirectional operation.} \label{fig3}
%\end{figure}

\begin{figure}[t]
\begin{center}
\includegraphics*[width=1\columnwidth,height=1.1in]{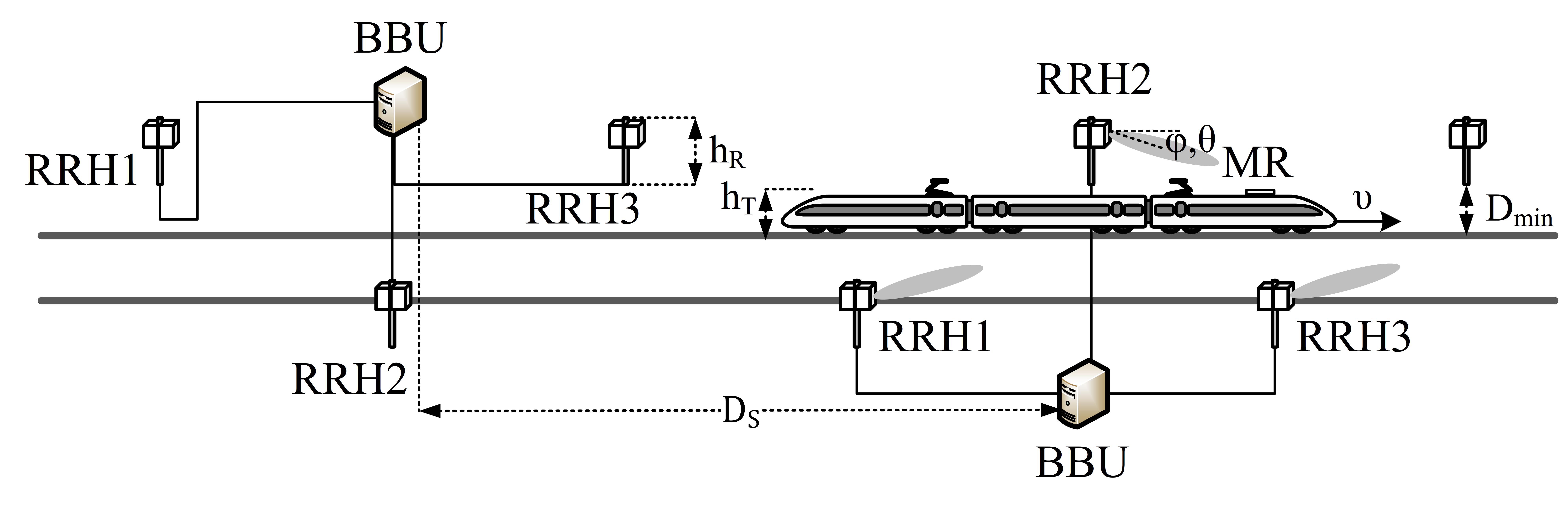}
\end{center}
\caption{Unidirectional HST SFN scenario.} \label{fig5}
\end{figure}

\begin{figure}[t]
\begin{center}
\includegraphics*[width=1\columnwidth,height=1.1in]{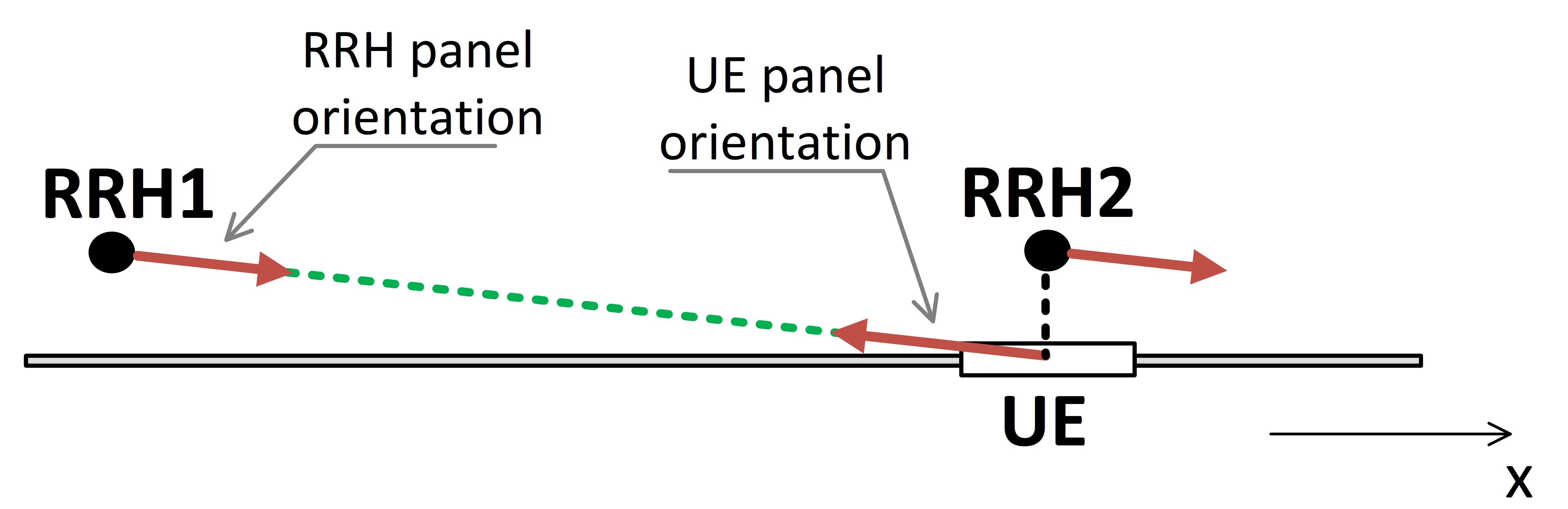}
\end{center}
\caption{RRH and UE panels orientation.} \label{fig6}
\end{figure}

Before introducing the mm-wave train-ground communication system model researched in this paper, we need to explain the concept of the single frequency network (SFN), which refers to a communication device deployment method where multiple remote radio heads (RRH) send and receive signals at the same frequency within a cell. When using this scheme, there is less interference from neighboring cells, cell coverage can be larger, and the frequency of handover is reduced accordingly. Therefore, this scheme is more suitable for HSR scenarios.

On the other hand, when studying the mm-wave train-ground communications, the main difference from the related research in the sub-6GHz frequency band is that we can no longer assume that the UE or RRH is equipped with an omnidirectional antenna. This is because the path loss of the mm-wave is larger than that of microwave, it is generally considered that the mm-wave communications are directional. Based on the above point of view, there are two types of SFN in HSR scenarios, which are unidirectional coverage and bidirectional coverage. In the unidirectional coverage scenario shown in Fig.~\ref{fig1}(a), the beams of the RRHs are always located on their left side, and the beam of the UE is located on its right side. Therefore, the UE always establishes a link with the RRH on its right in this scenario. In Fig.~\ref{fig1}(b), the UE's beam is still in unidirectional mode, but the RRHs can form beams on both sides. In this scenario, we also call the network working in the bidirectional coverage mode. In Fig~\ref{fig1}(c), both RRH and UE can form beams on both sides of themselves.

In this paper, we mainly consider the HSR SFN with unidirectional coverage, which corresponds to the scenario in Fig.~\ref{fig1}(a). The purpose of this is to provide a stable downlink carrier frequency, and this can be achieved by arranging the RRHs in such a manner that the strongest signal received by the UE has a nearly constant Doppler shift without sign-alternation. A stable downlink frequency, as experienced by the UE, leads to that uplink transmissions from the same UE are received by the RRHs with a nearly constant frequency offset~\cite{r1}.

The model of the mm-wave train-ground communication system researched in this paper is shown in Fig.~\ref{fig5}, which is the only one considered for HSR 30GHz deployment in 3GPP~\cite{r2}. In this model, a MR is deployed on the rooftop of the HST to overcome the huge penetration loss caused by the train body~\cite{r13}, ${D_{\min }}$ is distance between RRH and rail, ${D_s}$ is distance between adjacent RRH, ${h_R}$ is RRH height refer to rail, and ${h_T}$ is MR height, i.e., top of train. Three RRHs are connected over the fiber to the same base-band unit (BBU). In this paper, we consider that the RRH panel boresight direction points to the railway near the neighbouring RRH and the MR panel boresight direction has the correspondingly reciprocal angle as it is demonstrated in Fig.~\ref{fig6}.

%---------------------------
\subsection{Problem Formulation}\label{S3-2}
%---------------------------

In this paper, we will propose a DRL algorithm to fully learn the pattern of the mm-wave train-ground communication system to maximize the average RSP of the UE (i.e., MR) when there is no data available for advance training, and to further establish the mapping between position of the MR and the corresponding best beam direction.

According to the purpose of this paper and the calculation formula of link budget, when MR is located at $x$, the RSP, i.e., ${P_r}$ can be expressed as
\begin{align}
{P_r}(x,{\theta _E},{\phi _E},\theta _B^r,\phi _B^r) &\; = {P_t} + A_E^t({\theta _E},{\phi _E}) + A_E^r({\theta _E},{\phi _E})
\nonumber \\
&\; + A_B^t({\theta _E},{\phi _E}) - PL(x)
\nonumber \\
&\; + A_B^r({\theta _E},{\phi _E},\theta _B^r,\phi _B^r),
\label{eq1}
\end{align}
where ${P_t}$ is the transmit power of RRH. It should be noted here that the beamforming gain in the mm-wave communications is divided into the antenna element gain ${A_E}$ and the composite array radiation gain ${A_B}$, and so $A_E^t$ is the gain of the TX antenna element from RRH, and $A_E^r$ is the counterpart from MR. They are related to position $x$ of the MR, as well as the down-tilting angle ${\theta _E}$ and azimuth angle ${\phi _E}$ of the vector linked by the antenna element of MR and RRH. $A_B^t$ and $A_B^r$ are the composite array radiation gain from the RRH and MR, respectively. In this paper, we focus on the RX beamforming, so only consider adjusting the beam direction of the MR, i.e., the $\theta _B^r$ and $\phi _B^r$. In the subsequent simulations, the beam direction and width of RRH is fixed and not used as a research variable, so it is not given in~\eqref{eq1}.

On the other hand, it is unrealistic to adjust the beam direction at every position on the railway, and so we can consider fill the railway with location bins $[{d} - {\sigma _D},{d} + {\sigma _D}]$, and the beam direction of MR is adjusted only once per $2{\sigma _D}$. The introduction of location bin helps dynamically adjust the requirements for the accuracy of train location information in theoretical calculations and engineering practice. After that, whenever the MR enters a new location bin, the corresponding RSP needs to be calculated once. When the train reaches the end of the rail, all the obtained RSP values are arranged in a vector ${P_R}$ as shown in~\eqref{eq2}. The dimension $N$ of ${P_R}$ is the number of location bins on the rail, which can be expressed as~\eqref{eq3}.
\begin{equation}
{P_R} = [{P_r}({l_1}),{P_r}({l_2}),...,{P_r}({l_N})],
\label{eq2}
\end{equation}
\begin{equation}
N = \frac{L}{{2{\sigma _D}}} + 1.
\label{eq3}
\end{equation}

\begin{figure}[!t]
\begin{center}
\includegraphics*[width=0.7\columnwidth,height=1.2in]{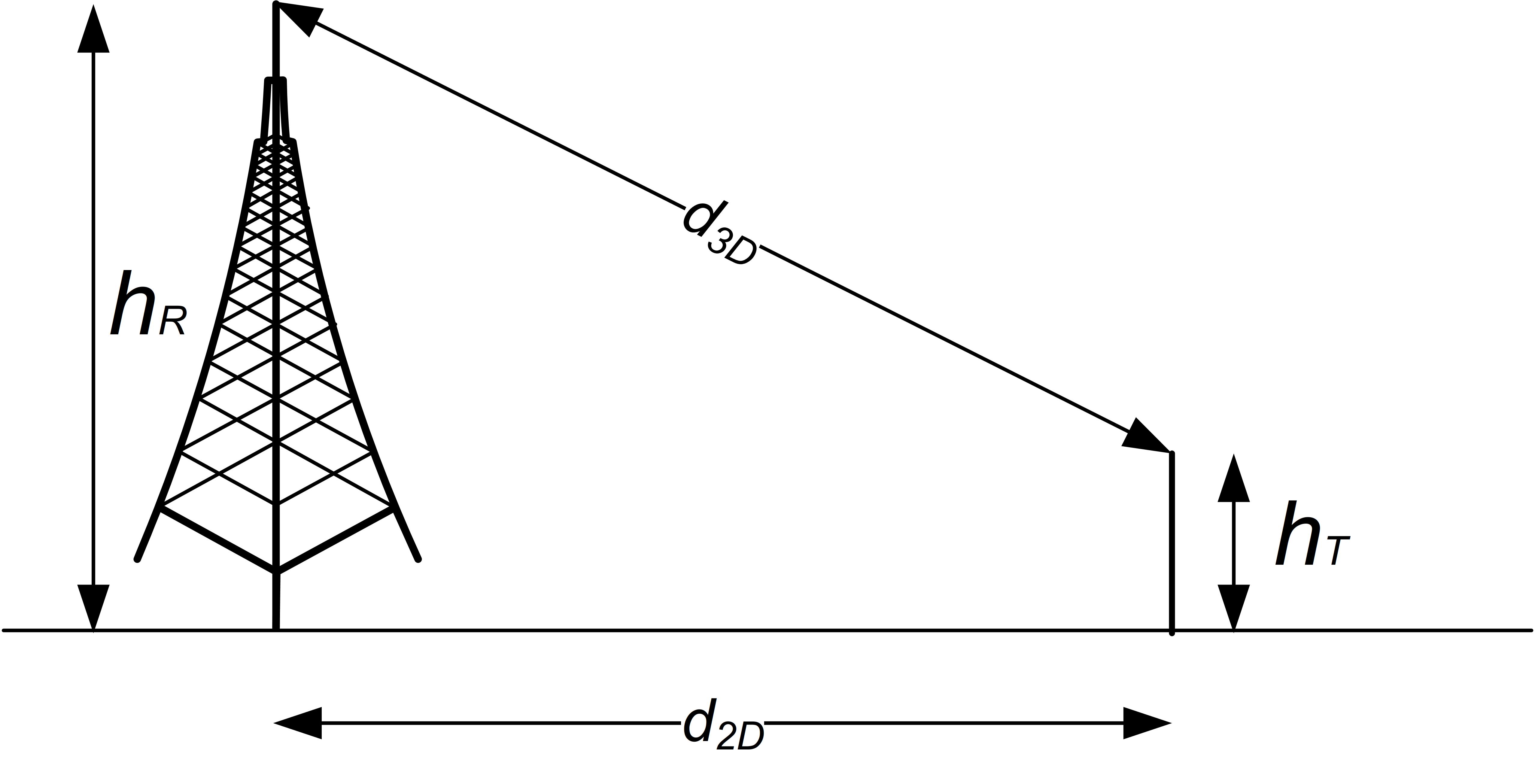}
\end{center}
\caption{Definition of ${d_{{\rm{2D}}}}$ and ${d_{{\rm{3D}}}}$.} \label{fig7}
\end{figure}

According to~\eqref{eq1}, the calculation of the RSP mainly includes three parts, namely path loss (PL), antenna element gain and composite array radiation gain. First, since our research scenario is HSR, considering that the HST mainly runs in the suburbs or viaducts, the signal is rarely blocked by buildings and causes non-line-of-sight (NLOS) transmission, and therefore only line-of-sight (LOS) transmission is considered in this paper. We use the large-scale PL model~\cite{r3} as shown in~\eqref{eq4}, ~\eqref{eq5} and ~\eqref{eq6}.
\begin{align}
& PL = \left\{ {\begin{array}{*{20}{c}}
{P{L_1},~~~~~{\rm{10m}} \le {d_{{\rm{2D}}}} \le {d_{{\rm{BP}}}}},\\
{{\rm{   }}P{L_2},~~~~{d_{{\rm{BP}}}} \le {d_{{\rm{2D}}}} \le 10{\rm{km}}},
\end{array}} \right.\label{eq4}\\
& P{L_1} = 20{\log _{10}}(40\pi  \cdot {d_{{\rm{3D}}}} \cdot \frac{{{f_c}}}{3})
\nonumber \\
&~~~~~~ - \min (0.044{h^{1.72}},14.77)
\nonumber \\
&~~~~~~ +  \min (0.03{h^{1.72}},10) \cdot {\log _{10}}({d_{{\rm{3D}}}})
\nonumber \\
&~~~~~~ + 0.002{\log _{10}}(h) \cdot {d_{{\rm{3D}}}},\label{eq5}\\
& P{L_2} = P{L_1}({d_{{\rm{BP}}}}) + 40{\log _{10}}\left(\frac{{{d_{{\rm{3D}}}}}}{{{d_{{\rm{BP}}}}}}\right),
\label{eq6}
\end{align}
where the definitions of ${d_{{\rm{2D}}}}$ and ${d_{{\rm{3D}}}}$ are shown in Fig.~\ref{fig7}. $h \in [5,50]{\rm{m}}$ represents the average height of buildings in the propagation environment. ${d_{{\rm{BP}}}}$ is the break point distance, and its calculation can be expressed as
\begin{equation}
{d_{{\rm{BP}}}} = 2\pi  \cdot {h_{\rm{R}}} \cdot {h_{\rm{T}}} \cdot \frac{{{f_c}}}{{\rm{c}}},
\label{eq7}
\end{equation}
where ${f_c}$ is the center frequency in Hz, and ${\rm{c}} = 3.0 \times {10^8}m/s$ is the propagation speed of electromagnetic waves in free space. It should be noted that ${f_c}$ in~\eqref{eq5} denotes the center frequency normalized by 1GHz.

In this paper, our purpose is to optimize the RX beamforming, thereby maximizing the average RSP of MR. Therefore, the gain from the antenna element and composite array radiation is the most important part. In this paper, the calculation of antenna element gain~\cite{r3} can be expressed as~\eqref{eq8},~\eqref{eq9} and~\eqref{eq10}.

\begin{small}
\begin{align}
& {A_E}\left( {{\theta'_E},\;{\phi'_E} = 0^\circ } \right) =  - \min \left\{ {12{{\left( {\frac{{{\theta'_E} - 90^\circ }}{{{\theta _{{\rm{3dB}}}}}}} \right)}^2},SL{A_V}} \right\},\label{eq8}\\
& {A_E}\left( {{\theta'_E} = 90^\circ ,\;{\phi'_E}} \right) =  - \min \left\{ {12{{\left( {\frac{{{\phi'_E}}}{{{\phi _{{\rm{3dB}}}}}}} \right)}^2},{A_{\max }}} \right\}.\label{eq9}
\end{align}
\end{small}

\begin{figure*}[bp]
\hrule
\begin{align}
{A_E}({\theta'_E},{\phi'_E}) =  - \min \left\{ { - \left( {{A_E}\left( {{\theta'_E},\;{\phi'_E} = 0^\circ } \right) + {A_E}\left( {{\theta'_E} = 90^\circ ,\;{\phi'_E}} \right)} \right),{A_{\max }}} \right\}.\label{eq10}
\end{align}
\end{figure*}

In~\eqref{eq8} and~\eqref{eq9}, ${\theta _{{\rm{3dB}}}}$ is the vertical 3 dB beamwidth, and ${\phi _{{\rm{3dB}}}}$ is the horizontal 3 dB beamwidth. $SL{A_V}$ stands for side-lobe attenuation in vertical direction, ${A_{\max }}$ is maximum attenuation. ${A_E}\left( {{\theta'_E},\;{\phi'_E} = 0^\circ } \right)$ and ${A_E}\left( {{\theta'_E} = 90^\circ ,\;{\phi'_E}} \right)$ are vertical and horizontal cut of the radiation power pattern in dB, respectively. It should be noted that ${\theta'_E}$ is the result of converting ${\theta_E}$ from the local coordinate system (LCS) to the global coordinate system (GCS), and the relationship between ${\phi'_E}$ and ${\phi _E}$ is the same.

On the other hand, the gain of the composite array radiation ${A_B}\left( {\theta ,\phi } \right)$ can be expressed as
\begin{small}
\begin{equation}
{A_B}\left( {{\theta _E},{\phi _E},{\theta _B},{\phi _B}} \right) = 10{\log _{10}}\left( {{{\left| {\sum\limits_{m = 1}^{{N_H}} {} \sum\limits_{n = 1}^{{N_V}} {{w_{n,m}} \cdot {v_{n,m}}} } \right|}^2}} \right),
\label{eq11}
\end{equation}
\end{small}where ${N_H}$ is the number of antenna elements on the panel, ${N_V}$ is the number of antenna elements with the same polarization in each column. ${v_{n,m}}$ is called the super position vector, ${w_{n,m}}$ is the weighting, and they can be expressed as~\eqref{eq12} and~\eqref{eq13},
\begin{figure*}[ht]
\begin{align}
& {v_{n,m}} = \exp \left( {i \cdot 2\pi \left( {\left( {n - 1} \right) \cdot \frac{{{d_V}}}{\lambda } \cdot \cos \left( {{\theta _E}} \right) + (m - 1) \cdot \frac{{{d_H}}}{\lambda } \cdot \sin ({\theta _E}) \cdot \sin ({\phi _E})} \right)} \right),
\label{eq12}\\
& {w_{n,m}} = \frac{1}{{\sqrt {{N_H}{N_V}} }}\exp \left( {i \cdot 2\pi \left( {\left( {n - 1} \right) \cdot \frac{{{d_V}}}{\lambda } \cdot \sin \left( {{\theta _B}} \right) - (m - 1) \cdot \frac{{{d_H}}}{\lambda } \cdot \cos ({\theta _B}) \cdot \sin ({\phi _B})} \right)} \right).
\label{eq13}
\end{align}
\hrule
\end{figure*}
where $i$ represents the imaginary unit, $\lambda$ is the wavelength of the carrier, and ${d_V}$ represents the distance of the antenna element in the vertical direction, and generally set ${d_V} = \frac{\lambda }{2}$.

It should be noted, taking into account the difference in physical characteristics of the RRH and MR antennas in practice, the obtainment of antenna gain should be divided into two parts: the MR side and RRH side. The values of the parameters in the simulation will be given in Section~\ref{S5-1}.

Due to the short wavelength, mm-wave is vulnerable to various blockages, such as foliage, buildings, and viaducts in HSR scenarios. The blockage occurs when obstacles appear in the radio links between the transceivers, which results in RSP degradation caused by severe attenuation~\cite{r31}. When the achieved signal to interference plus noise ratio (SINR) at the RX side is lower than the required threshold ${\gamma _{th}}$, the system is unable to guarantee the required bit error rate and the link is considered to be turned off~\cite{r32}. The link blockage depends on multiple factors, including the surrounding environment, obstacle density, beamwidth, and transmission distance.

In this paper, we focus on RX beamforming in mm-wave train-ground communication system. For the purpose of simplifying the analysis, we previously assume that only one MR is set on the train roof. In order to better deal with the blockages, we will discuss this problem under the assumption of setting multiple MRs. Considering the height of the MRs and HSR mainly operates in rural areas and on viaducts, the link between the MRs and the BS can be regarded as LOS transmission. We consider that the blockage problem in the investigated system can be solved in three steps. First, we define an average outage probability ${P_b}$ for all links, which note that each link has a random probability of experiencing blockage. Secondly, if the link between an MR and the BS is blocked, the users associated with the MR will re-establish a connection with the next nearest MR to avoid the blockage. Third, we study the scenario of NLOS transmission, which can be enhanced with the assist of intelligent reflecting surface (IRS) and Device-to-Device (D2D) communications. In this paper, for analytical simplicity, it is assumed that the link blockage probability remains stable in a road segment. The link blockage probability ${P_b}$ on average is deemed as a constant in a certain section along the rail track~\cite{r33}.

In summary, the objective function and constraints of the RSP maximization problem researched in this paper can be expressed as
\begin{align}
%\begin{array}{l}
\mbox{(P1)} \;\;
\max &~~{\rm{ }}(1-P_b){\left\| {{P_R}} \right\|_2}{\rm{   }} \label{P1}
\\
\mbox{s.t.}&~~{P_r}(l) \le {P_t}(l),{\rm{  }}l = {l_1},{l_2},...{l_N}, \label{P1-c1} \\
&~~{\theta _B} \in [{0^ \circ },{180^ \circ }],\label{P1-theta}\\
&~~{\phi _B} \in [ - {180^ \circ },{180^ \circ }],\label{P1-phi}\\
&~~{\rm{A}}_E^t \le {\rm{A}}_{E,\max }^t, \label{P1-c2}\\
&~~{\rm{A}}_E^r \le {\rm{A}}_{E,\max }^r, \label{P1-c3}
%\end{array}
\end{align}
where ${P_t}(l)$ and ${P_r}(l)$ correspond to the transmitted and received signal power at location bin $l$, respectively.
\eqref{P1-theta} and~\eqref{P1-phi} specify the search range of the beam direction.
${\rm{A}}_{E,\max }^t$ and ${\rm{A}}_{E,\max }^r$ are the maximum directional gain from the RRH and MR, respectively.

Obviously, P1 is a mixed integer nonlinear programming problem, usually NP-hard. We hope to propose an efficient DRL algorithm to optimize the beam directions ${\theta _B}$ and ${\phi _B}$ of the MR, so that the maximum average RSP can be obtained.

Based on the above mathematical relationship, if don't considering the correlation between different location bins when calculating the RSP, we can obtain separately the maximum RSP available in each location bin, and the solution of P1 can be obtained by traversing each location bin. But in practice, the HST runs continuously on the railway, and so the optimal beam direction in the adjacent position is obviously related. However, this relationship is not easy to describe with a mathematical model at present. Traditional beam management methods, such as EBS, ignore this correlation, and so the corresponding overhead and delay are high. Because the LOS component in the researched mm-wave train-ground communication system is dominant, the directional characteristics are more protruding, and the correlation between the position of MR and its optimal beam direction is also stronger.

According to the formulated objective function, the optimization goal of this paper is to maximize the average RSP of MR over a long-distance rail, so as to explore the wireless coverage and equipment deployment of the mm-wave train-ground communication system in future work, which will do a good job in technical research and standardization for the landing of FR2 HST communications. Therefore, we paid more attention to the impact of large-scale fading.

In order to effectively solve P1, we will introduce a DRL algorithm to establish the mapping between the position of MR and its optimal beam direction, as well as the optimal beam direction and the maximum average RSP in the next section.

%---------------------------
\section{RX Beamforming Optimization Via Deep Reinforcement Learning}\label{S4}
%---------------------------
%Based on a fixed timetable, HST runs periodically on the same railway, and so the train-ground communication system also accordingly has regular patterns in time and space. The correlation between the position of MR and the its optimal beam direction, as well as the optimal beam direction and the maximum average RSP, can be considered as a pattern of the mm-wave train-ground communication system in space. At present, because artificial intelligence (AI) algorithms are good at extracting hidden relationships between data,

In this paper, we focus on improving the RSP of MR in the investigated mm-wave train-ground communication system from the perspective of RX beamforming. Scholars often face the trade-off between overhead and system performance in the existing related work. For example, compared with beam switching, beam tracking can achieve higher antenna gain, but brings higher computational complexity, and traditional optimization algorithms that require massive iterations are obviously not suitable for solving communication problems in HSR scenarios. At the same time, beam switching is achieved by sacrificing certain system performance in exchange for improved complexity.

On the other hand, the route of the HST is fixed and its operation has a certain periodicity. Therefore, the track-side train-ground wireless communication system should have some regular patterns, and the correlation between the position of MR and the optimal beam direction, as well as the optimal beam direction and the maximum average RSP, can be considered as a pattern of the train-ground communication system in space, and artificial intelligence (AI) is good at capturing these hidden patterns. However, the research on using AI to capture hidden patterns in mm-wave train-ground communications is still scarce, so we can consider design efficient solutions from this point. In addition, compared with other AI algorithms, DRL does not require a large number of training samples and environmental information, and its computational complexity is mainly in the offline training phase. In the online stage, the most valuable action will be executed directly after the agent observes the state of the system. Because of these advantages, the application of DRL in wireless communication has attracted more and more attention from scholars and industry~\cite{r6}. Therefore, DRL has the potential to help us improve the system performance while ensuring low complexity.

%Compared with other AI algorithms, RL is model-free and doesn't require sample data from external supervisors~\cite{r5}.

The proposed solution in this paper for solving P1 consists of two parts: Section \ref{S4-1} is a Q-learning algorithm for online adjustment of beam direction, and Section \ref{S4-2} is a convolutional neural network (CNN), which is a variant of a deep neural networks (DNN) and used to estimate the Q-values offline.

%---------------------------
\subsection{RX Beamforming Based on Reinforcement Learning}\label{S4-1}
%---------------------------

RL is a new paradigm for intelligent decision-making, and can be implemented with the help of TensorFlow and Keras. The basic idea of RL is to set a agent to interact with the environment and use the training experience to learn the Markov decision process model~\cite{r4}.

In the established system model, we consider a RL agent deployed in the cloud whose responsibility is to train and output the best strategy for adjusting the beam direction when maximizing average RSP is the optimization target. In particular, this agent will interact with the mm-wave train-ground communication system for a long time. It observes the state of the system and finds the cumulative reward that can be obtained. In the decision-making stage, the agent derives a control strategy and specifies an action. After executing the selected action, the system will enter a new state, and then the agent continues to make new decisions.

In order to obtain the optimal control strategy, first, we need to clarify the system state space, action space and reward function as follows:

%---------------------------
\subsubsection{System State Space}\label{S4-1-1}
%---------------------------

One of the purposes of this paper is to establish a mapping between the position of the MR and its optimal beam direction. Considering the railway as the x-axis, the state is defined as the position of the MR projected on the x-axis, i.e., the position of the MR relative to the rail. When the length of a location bin is $2{\sigma _D}$ and the railway is $L$, the state space of the system is $S = ({s_1},{s_2},...,{s_N})$, where $N$ can be obtained by~\eqref{eq3}, ${s_n} = (n - 1) \cdot 2{\sigma _D}$, $n = 1,2,...N$.

%---------------------------
\subsubsection{Action Space}\label{S4-1-2}
%---------------------------

Considering that there is a correlation between the optimal beam directions when MR is located in adjacent loaction bins, and the adjustment of antenna arrays has the physical continuity, we can establish a relationship that the optimal beam direction corresponding to the latter location bin is related to the counterpart of the previous location bin. Therefore, we define the action space as $A = ({a_1},{a_2},...{a_9})$. When MR is located in the  $k$-th $(k \ge 1)$ location bin ${l_k}$, the optional actions are shown in Table I, where $\theta _B^{k - 1}$ and $\phi _B^{k - 1}$ are the beam angles after adjustment when the MR is located in the $(k - 1)$-th location bin, and ${\sigma _B}$ is the step size of angle adjustment. In addition, both $\theta _B^0$ and $\phi _B^0$ are defined as 0$^\circ $.

%\begin{table}[!t]
%	\begin{center}
%		\caption{Action Space and Its Definition}
%		\begin{tabular}{c|c|c|c}
%			\toprule
%			Action & Definition \\
%			\midrule
%			${a_1}$ & $\theta _B^k = \theta _B^{k - 1},\phi _B^k = \phi _B^{k - 1}$ \\
%			${a_2}$ & $\theta _B^k = \theta _B^{k - 1},\phi _B^k = \phi _B^{k - 1} + {\sigma _B}$ \\
%			${a_3}$ & $\theta _B^k = \theta _B^{k - 1} + {\sigma _B},\phi _B^k = \phi _B^{k - 1}$ \\
%			${a_4}$ & $\theta _B^k = \theta _B^{k - 1} + {\sigma _B},\phi _B^k = \phi _B^{k - 1} + {\sigma _B}$ \\
%			${a_5}$ & $\theta _B^k = \theta _B^{k - 1},\phi _B^k = \phi _B^{k - 1} - {\sigma _B}$ \\
%			${a_6}$ & $\theta _B^k = \theta _B^{k - 1} - {\sigma _B},\phi _B^k = \phi _B^{k - 1}$ \\
%			${a_7}$ & $\theta _B^k = \theta _B^{k - 1} - {\sigma _B},\phi _B^k = \phi _B^{k - 1} - {\sigma _B}$ \\
%			${a_8}$ & $\theta _B^k = \theta _B^{k - 1} + {\sigma _B},\phi _B^k = \phi _B^{k - 1} - {\sigma _B}$ \\
%			${a_9}$ & $\theta _B^k = \theta _B^{k - 1} - {\sigma _B},\phi _B^k = \phi _B^{k - 1} + {\sigma _B}$ \\
%			\bottomrule
%		\end{tabular}
%		\label{table1}
%	\end{center}
%\end{table}

\renewcommand\arraystretch{1.5}
\begin{table}[!t]
	\begin{center}
		\caption{Action space and its definition}
		\begin{tabular}{c|c|c}
			\toprule
			Action & \multicolumn{2}{c}{Definition} \\
			\midrule
			~ & Down-tilting angle & Azimuth angle \\
			\midrule
			${a_1}$ & $\theta _B^k = \theta _B^{k - 1}$ & $\phi _B^k = \phi _B^{k - 1}$ \\
			${a_2}$ & $\theta _B^k = \theta _B^{k - 1}$ & $\phi _B^k = \phi _B^{k - 1} + {\sigma _B}$ \\
			${a_3}$ & $\theta _B^k = \theta _B^{k - 1} + {\sigma _B}$ & $\phi _B^k = \phi _B^{k - 1}$ \\
			${a_4}$ & $\theta _B^k = \theta _B^{k - 1} + {\sigma _B}$ & $\phi _B^k = \phi _B^{k - 1} + {\sigma _B}$ \\
			${a_5}$ & $\theta _B^k = \theta _B^{k - 1}$ & $\phi _B^k = \phi _B^{k - 1} - {\sigma _B}$ \\
			${a_6}$ & $\theta _B^k = \theta _B^{k - 1} - {\sigma _B}$ & $\phi _B^k = \phi _B^{k - 1}$ \\
			${a_7}$ & $\theta _B^k = \theta _B^{k - 1} - {\sigma _B}$ & $\phi _B^k = \phi _B^{k - 1} - {\sigma _B}$ \\
			${a_8}$ & $\theta _B^k = \theta _B^{k - 1} + {\sigma _B}$ & $\phi _B^k = \phi _B^{k - 1} - {\sigma _B}$ \\
			${a_9}$ & $\theta _B^k = \theta _B^{k - 1} - {\sigma _B}$ & $\phi _B^k = \phi _B^{k - 1} + {\sigma _B}$ \\
			\bottomrule
		\end{tabular}
		\label{table1}
	\end{center}
\end{table}

%---------------------------
\subsubsection{Reward Function}\label{S4-1-3}
%---------------------------

Before introducing the reward function, we need to define a benchmark scheme in which the beam direction of MR are constant at ${\theta _F}$ and ${\phi _F}$. Then, the reward function can be expressed as
\begin{equation}
R({s^k},{a^k}) = {P_r}({s^k},{a^k}) - {P_r}({s^k},{\theta _F},{\phi _F}),
\label{eq15}
\end{equation}
where ${P_r}({s^k},{a^k})$ represents the RSP of MR after the selected action is executed in the system state ${s^k}$ of the $k$-th epoch, and ${P_r}({s^k},{\theta _F},{\phi _F})$ is the RSP when the benchmark scheme is adopted in the same system state.

Due to the dynamic nature of the HSR scenario, the state of the train-ground communication system changes with the passage of time and the displacement of MR. In this process, the agent continuously interacts with this communication system. In the $k$-th epoch, the agent obtains state ${s^k}$ of the system, i.e., the position of the MR, and then derives an optimal strategy $\pi$ through the learning. After that, the system executes action ${a^k}$ of adjusting the beam direction based on strategy $\pi$, and obtains reward $R({s^k},{a^k})$ according to the reward function. Finally, the system advances to a new state ${s^{k + 1}}$, and the agent continues to repeat the above operations until it reaches the last learning cycle. In addition, the reward returned by the algorithm is the accumulated and discounted reward, which is used to measure how well action ${a^k}$ is executed under state ${s^k}$, and can be expressed as
\begin{equation}
{R_k} = \sum\limits_{t = 0}^T {\alpha R({s^{k + t}},{a^{k + t}})},
\label{eq16}
\end{equation}
where $t$ is the epoch, $T$ is the maximum learning period, and $\alpha  \in (0,1]$ is the discount factor.

It should be noted here that in the process of RL, the goal of the agent is to maximize the expected accumulated reward, which is $\max E[{R_k}|{s^k}]$. Therefore, we not only concern about the instant reward obtained after executing action in the current state, but also hope that the cumulative discounted rewards will be as large as possible in the long term. In order to obtain $\max E[{R_k}|{s^k}]$, researchers have proposed two widely used methods: the value function-based~\cite{r7} and the strategy-based~\cite{r8}. In this paper, we adopt the value function-based method.

The value function is a essential element in RL and Q-learning is a widely used algorithm for obtaining the value function of state-action pairs. In order to determine the value function, we first need to define a state value function, represented by ${V_\pi }({s^k})$, which is the cumulative reward of strategy $\pi$. Therefore, we have ${V_\pi }({s^k}) = E[{R_k}|{s^k}]$. Considering that under the researched mm-wave train-ground communication system, the system state changes are independent, we can rewrite the state value function as follows:
\begin{equation}
{V_\pi }({s^k}) = r({s^k},{\pi _k}) + \alpha \sum\limits_{{s^{k'}} \in S} {{P_{{s^k}{s^{k'}}}}({\pi _k}){V_\pi }({s^{k'}})},
\label{eq17}
\end{equation}
where ${P_{{s^k}{s^{k'}}}}({\pi _k})$ represents the state transition probability when strategy ${\pi _k}$ is adopted.

Next, we can define a Q-function ${Q_\pi }({s^k},{a^k})$ representing the value of the state-action pairs to describe the expected reward after the action ${a^k}$ is executed when strategy $\pi$ is adopted and the system is in state ${s^k}$. Therefore, we have ${Q_\pi }({s^k},{a^k}) = E[{R_k}|{s^k},{a^k}]$. Since the system state, i.e., the change of the MR position is independent of the adjustment of the beam direction, we have ${Q_\pi }({s^k},{a^k}) = {V_\pi }({s^k})$, which can be expressed as
\begin{equation}
{Q_\pi }({s^k},{a^k}) = r({s^k},{a^k}) + \alpha \sum\limits_{{s^{k'}} \in S} {{P_{{s^k}{s^{k'}}}}({a^k}){V_\pi }({s^{k'}},{a^{k'}})}.
\label{eq18}
\end{equation}

At this time, the goal of the agent is to maximize this Q-function, so as to obtain the optimal strategy for maximizing the average RSP researched in this paper.

What we need to note is that $r({s^k},{a^k})$ and ${P_{{s^k}{s^{k'}}}}({a^k})$ are unknown in P1. In order to solve this problem, Q-learning is one of the most commonly used algorithms. In Q-learning, the Q-function can usually be determined recursively, and its update can be expressed as
\begin{align}
Q({s^{k + 1}},{a^{k + 1}}) & = Q({s^k},{a^k}) + \beta  \cdot r({s^k},{a^k})
\nonumber \\
& + \alpha [\mathop {\max }\limits_{{{a'}^k}} Q({s'^k},{a'^k})] - Q({s^k},{a^k}),
\label{eq19}
\end{align}
where $\beta $ represents the learning rate, ${s'^k}$ and ${a'^k}$ represent the state in the system state space and action in the action space at epoch $k$, respectively.

Obviously, according to~\eqref{eq19}, each state-action pair needs to be accessed and evaluated when updating $Q({s^{k + 1}},{a^{k + 1}})$, which leads to huge complexity and slow convergence. Therefore, when the problem to be solved is a large-scale optimization problem, the actual performance of this RL algorithm is not good. Therefore, in order to make the algorithm have better performance for solving P1, we will further introduce DNN to achieve more intelligent control.

\subsection{RX beamforming Based on DQN}\label{S4-2}
In order to accelerate the training of the Q-learning, we propose an improved RL algorithm which is DQN to further solve P1. This method uses sampled data to train a NN for estimating Q-values, which can map the inputs of state-action pairs to their corresponding Q-values~\cite{r9}. However, due to the correlation between the training samples and the correlation between Q-values and target values, applying the NN directly to Q-learning may cause the results to fail to converge~\cite{r4}.
\begin{figure}[!t]
\begin{center}
\includegraphics*[width=1\columnwidth,height=2.3in]{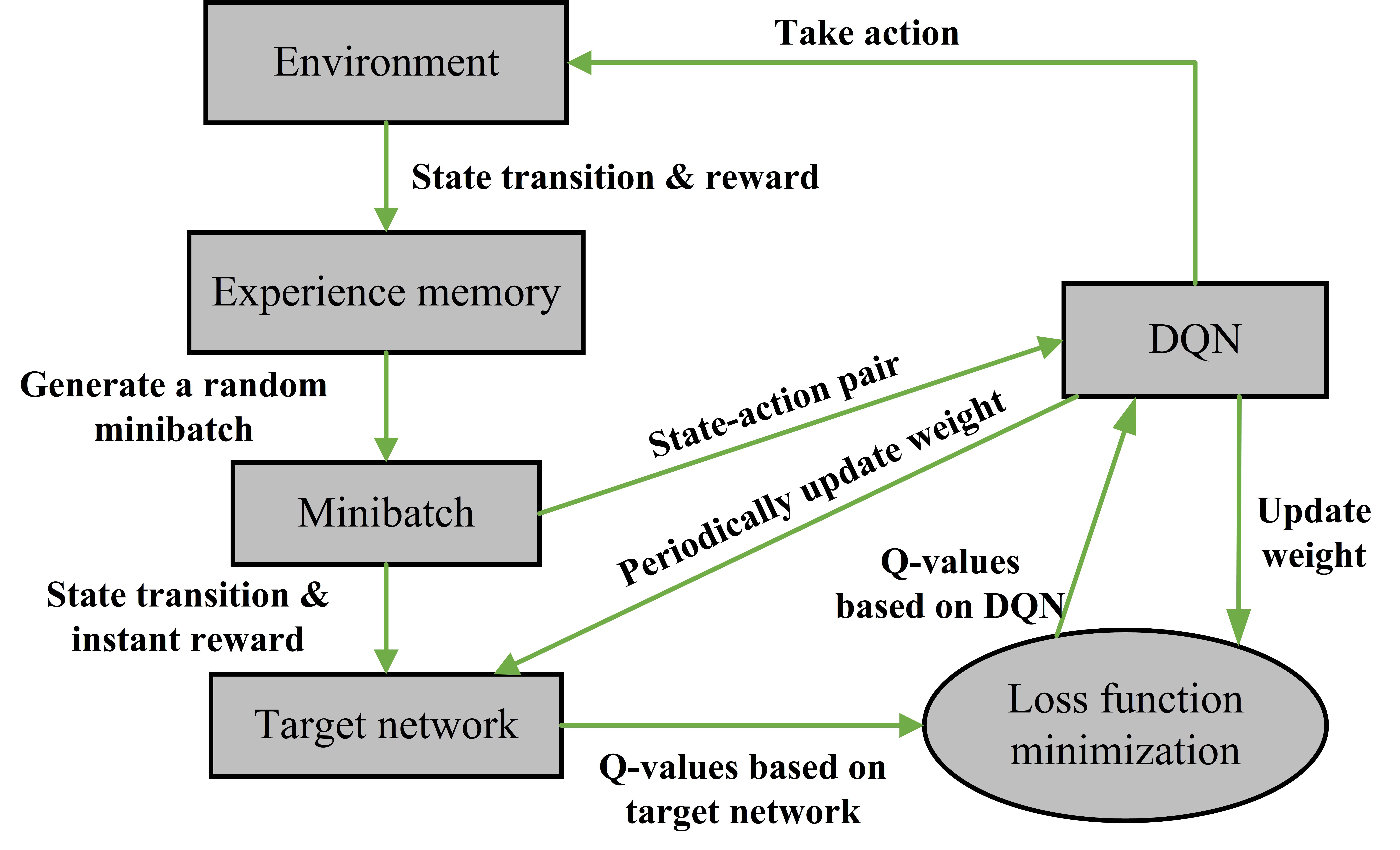}
\end{center}
\caption{Frame of the DRL approach.} \label{fig8}
\end{figure}

In order to reduce this correlation, the DRL method has been proposed~\cite{r6}, in which a DNN is used to estimate Q-values to generate a DQN. In this DRL algorithm, the agent first explores the environment by executing actions randomly, and stores the experience in the target network. A set of experience includes the current system state, executed action, instant reward and new state. Then a mechanism called experience replay is used, where the data are randomly sampled in minibatches from the target network to break the correlation in a sequence of observation. With samples from the target network, the weights of the DQN are updated by minimizing the mean square error of Q-functions between the DQN and the target network~\cite{r4}. We usually use the stochastic gradient descent method to update the weight parameters. The general framework of the DRL approach is shown in Fig.~\ref{fig8}, and the specific mathematical process is as follows.

First, we replace the value function ${Q_\pi }({s^k},{a^k})$ by a DQN with parameters ${\omega ^k}$, i.e., ${Q_\pi }({s^k},{a^k}) \approx Q({s^k},{a^k},{\omega ^k})$, and this approximation is used to define the loss function as shown in~\eqref{eq20}.
\begin{align}
\begin{array}{l}
L({\omega ^k}) = E[(\underbrace {r({s^k},{a^k}) + \beta  \cdot \mathop {\max }\limits_{{a^{k + 1}}} Q({s^{k + 1}},{a^{k + 1}},{\omega ^k})}_{{\rm{Target}}}\\
~~~~~~~~~~~~~~ - Q({s^k},{a^k},{\omega ^k}){)^2}].
\end{array}
\label{eq20}
\end{align}
Then the gradient of the loss function with respect to the parameter ${\omega ^k}$ is obtained, which can be expressed as
\begin{equation}
\begin{array}{l}
\frac{{\partial L({\omega ^k})}}{{\partial {\omega ^k}}} =  - E[(\underbrace {r({s^k},{a^k}) + \beta  \cdot \mathop {\max }\limits_{{a^{k + 1}}} Q({s^{k + 1}},{a^{k + 1}},{\omega ^k})}_{{\rm{Target}}}\\
~~~~~~~~~~~~~~~~~- Q({s^k},{a^k},{\omega ^k}))\frac{{\partial Q({s^k},{a^k},{\omega ^k})}}{{\partial {\omega ^k}}}].
\end{array}
\label{eq21}
\end{equation}

In each epoch, the algorithm repeat this process until reach the last learning cycle. It can update ${\omega ^k}$ based on the experience randomly selected from the experience pool to obtain the best strategy for maximizing the average RSP of the MR. After the DQN completes the training, the agent executes actions based on the estimated Q-values.

The pseudo code of the DQN algorithm for optimizing RX beamforming is given in Algorithm \ref{alg2}, where $K$ indicates the largest episode. In each episode, the proposed algorithm consists of two stages. The first is the transition generating stage, which uses the Q-learning to adjust the direction of the RX beam online. At this stage, the agent observes the state of the system, randomly selects action with probability $\varepsilon $ or that of the most likely to obtain the largest discount accumulative reward with probability $1 - \varepsilon $, and executes the selected action, gets the corresponding reward. Then the system reaches the next state, and stores this memory into ${\cal B}$ as experience. The second is the parameter updating stage. At this stage, the proposed algorithm randomly samples minibatch of memories from ${\cal B}$, and uses the stochastic gradient descent method to update the parameters of NN to make the prediction of Q-values more accurate.

\begin{algorithm}[t]
\caption{The DQN based algorithm for RX Beamforming}
\label{alg2}
%Acquire the mm-wave train-ground communication environment information.\\
%Initialize RX beamforming: $\theta _B^r({l_0}) = 0^\circ$, $\phi _B^r({l_0}) = 0^\circ$, $\theta _B^t(l) = \phi _B^t(l) = 0^\circ ,l = {l_1},{l_2},...,{l_N}$.\\
%Initialize action-value function $Q$, target action-value function $\hat Q$ and replay memory buffer ${\cal B}$.
\begin{algorithmic}[1]
\REQUIRE ~~\\
   Acquire the HSR environment information.\\
   Initialize RX beamforming: $\theta _B^r({l_0}) = 0^\circ$, $\phi _B^r({l_0}) = 0^\circ$, $\theta _B^t(l) = \phi _B^t(l) = 0^\circ ,l = {l_1},{l_2},...,{l_N}$.\\
   Initialize action-value function $Q$, target action-value function $\hat Q$ and replay memory buffer ${\cal B}$.
\FOR {${\rm{episode~ }}k = 1:K$}
\STATE Initialize state sequence $s_1^k = 0{\rm{m}}$.
\FOR {$n = 1:N$}
\STATE Observe current state $s_n^k = (n - 1) \cdot 2{\sigma _D}$.
\renewcommand{\algorithmicrequire}{ \textbf{In the First Stage:}}
\REQUIRE // Transition Generating Stage
\STATE Select an action $a_n^k$ form $A$ based on $\varepsilon  - greedy$ policy;
\STATE Execute the action $a_n^k$, receive a reward $R(s_n^k,a_n^k)$ and the next state $s_{n + 1}^k$;
\STATE Store $\{ s_n^k,a_n^k,R(s_n^k,a_n^k),s_{n + 1}^k\}$ into ${\cal B}$.
\renewcommand{\algorithmicrequire}{ \textbf{In the Second Stage:}}
\REQUIRE // Parameter Updating Stage
\STATE Sample random minibatch of experience $\{ {s_m},{a_m},R({s_m},{a_m}),{s_{m + 1}}\} $ from ${\cal B}$.
\IF{episode terminates at step $m+1$}
\STATE set ${y_m} = R({s_m},{a_m})$;
\ELSE
\STATE set ${y_m} = R({s_m},{a_m}) + \beta  \cdot \mathop {\max }\limits_{a'} \hat Q({s_{m + 1}},a')$.
\ENDIF
\STATE Perform a gradient descent step on ${({y_m} - Q({s_m},{a_m}))^2}$.
\STATE Every $C$ steps reset $\hat Q = Q$.
\ENDFOR
\ENDFOR
\renewcommand{\algorithmicrequire}{ \textbf{Output:}}
\REQUIRE~~\\
   RX power dictionary $\{ {P_R}\} _{n = 1}^N$.\\
   Well-trained parameters.
\end{algorithmic}
\end{algorithm}

Based on the obtained best strategy, we establish a mapping between the position of MR and its optimal beam direction, as well as that and the maximum average RSP. Then, we can consider establishing a database as shown in Table II.
\begin{table}[t]
	\begin{center}
		\caption{An example of the database generated after learning}
        \setlength\tabcolsep{3pt}{
		\begin{tabular}{|c|c|c|c|c|c|c|c|}
			\hline
			\multirow{2}{*}{Obsv.No.} & \multicolumn{2}{c|}{Best} & \multicolumn{2}{c|}{2nd best} & ... & \multicolumn{2}{c|}{${D_C}$-th best} \\
			\cline{2-8}
			& $[\theta _B^r,\phi _B^r]$ & ${P_r}$ & $[\theta _B^r,\phi _B^r]$ & ${P_r}$ & ... &  $[\theta _B^r,\phi _B^r]$ & ${P_r}$ \\
			\hline
			${l_1}$ & [17,30] & -80.2 & [20,30] & -83.1 & ... & [8,35] & -90.6 \\
			${l_1}$ & [23,30] & -85.3 & [20,30] & -85.7 & ... & [10,45] & -91.2 \\
			... & ... & ... & ... & ... & ... & ... & ... \\
			${l_N}$ & [85,50] & -101.8 & [79,50] & -102.2 & ... & [70,68] & -103.9 \\
			\hline
		\end{tabular}}
		\label{table2}
	\end{center}
\end{table}

According to Table II, based on the optimal strategy $\pi$ obtained by learning, the database contains the RSP values of ${D_C}$ candidate optimal beam directions in each location bin on the railway, which are arranged in each column in the database in order of magnitude, and the front beam direction corresponds to a better Q-value. When the MR enters a new location bin, the agent can choose ``utilization" or ``exploration". When it selects ``utilization" with probability ${P_1}$, the current beam direction of MR will be directly adjusted to the optimal beam direction by viewing the database. On the contrary, when the agent chooses ``exploration" with probability $1 - {P_1}$, the adjustment of the beam direction will no longer consider the optimal beam recorded in the database, but randomly select from other ${D_C} - 1$ beam directions. After ``utilization" or ``exploration", the new value of RSP will overwrite the corresponding data in the database as an update. The main purpose of introducing this database is to make the proposed solution more suitable for the mobile characteristics of the HSR scenario. In a changing communication environment, we hope that the proposed algorithm can be fine-tuned under the basic system model. At the same time, it should be noted that the purpose of our research is to use a DRL algorithm to optimize RX beamforming to maximize the average RSP in the mm-wave train-ground communication system, not to improve the performance of the DRL algorithm itself.

The computational complexity analysis of the proposed algorithm is crucial and necessary, which mainly includes two parts, the offline and the online training phase. The offline phase is composed of transition generating stage and parameter updating stage. Assuming that the number of epochs is $E$, and the computational complexity of calculating RSP is $T$, then the computational complexity of transition generating stage is ${\rm{O}}(ENT)$. The size of data set where state-action pairs are saved is marked as $D$, and the batch size as $B$, then the computational complexity of DQN training is ${\rm{O}}(E \times \frac{D}{B} \times T)$. In the online phase, the proposed algorithm first observes the state of the system, i.e., agent obtains the position information of the HST, and executes the most valuable action. Therefore, the computational complexity in the online phase is ${\rm{O}}(NT)$. In summary, Algorithm 1 yields the computational complexity of ${\rm{O}}(E(\frac{D}{B} + N)T + NT)$.

%---------------------------
\section{Performance Evaluation}\label{S5}
%---------------------------

%---------------------------
\subsection{Simulation Setup}\label{S5-1}
%---------------------------

According to the system model illustrated in Section~\ref{S3-1}, the train-ground communication system works at 30 GHz mm-wave frequency band. On the other hand, because mm-wave BSs are usually used in small-scale and densely deployed communication scenarios, and HST running in suburbs and viaducts helps the directional propagation of mm-wave, the length of the rail $L$ is set to 2,000m, and the RRH is located 700m from the beginning of the rail in the simulation.

In this paper, based on the DRL framework, designing an efficient RX beamforming scheme to improve the average RSP of MR is our research goal. Therefore, in order to eliminate the influence of BS and simplify the analysis, we use the widest possible transmit beam to ensure coverage. He \emph{et al.} mentioned that to guarantee communication links at large distance, 0 degree rotate angle is the optimum selection~\cite{He0}. Therefore, in the simulation, we set the azimuth and down-tilting of the transmitting beam to be $0^\circ$. Other simulations parameters are given in Table~\ref{table3}.

\renewcommand\arraystretch{1.2}
\begin{table}[t]
\begin{center}
\caption{Simulation Parameters}
\begin{tabular}{l|c}
\toprule
Parameters & Values \\
\midrule
Distance between RRH and rail: ${D_{\min }}$ & 150 m  \\
Distance between adjacent RRH: ${D_S}$ & 700 m \\
RRH height: ${D_{{\rm{RRH\_height}}}}$ & 15 m \\
MR height (top of train): ${D_{{\rm{MR\_height}}}}$  & 5 m \\
Transmit power: ${P_t}$ & 31 dBm \\
Down-tilting angle of the RRH beam: $\theta _B^t$ & $0^{\circ}$ \\
Azimuth angle of the RRH beam: $\phi _B^t$ & $0^{\circ}$ \\
Radius of a location bin: ${\sigma _D}$ & 2.5 m \\
Step size of beam angle adjustment: ${\sigma _B}$ & $3^{\circ}$ \\
The average height of the building: $h$ & 5 m \\
Number of BS antennas: $N_T$ & 8 \\
Number of MR antennas: $N_R$ & 8 \\
vertical 3dB beamwidth (RRH): ${\theta _{{\rm{RRH\_3dB}}}}$ & $65^{\circ}$ \\
vertical 3dB beamwidth (MR): ${\theta _{{\rm{MR\_3dB}}}}$ & $90^{\circ}$ \\
horizontal 3dB beamwidth (RRH): ${\phi _{{\rm{RRH\_3dB}}}}$ & $65^{\circ}$ \\
horizontal 3dB beamwidth (MR): ${\phi _{{\rm{MR\_3dB}}}}$ & $90^{\circ}$ \\
side-lobe attenuation in vertical direction (RRH): $SL{A_{{\rm{RRH\_}}V}}$ & 30 dB \\
side-lobe attenuation in vertical direction (MR): $SL{A_{{\rm{MR\_}}V}}$ & 25 dB \\
maximum attenuation (RRH): ${A_{{\rm{RRH\_}}\max }}$ & 30 dB \\
maximum attenuation (MR): ${A_{{\rm{MR\_}}\max }}$ & 25 dB \\
maximum directional gain (RRH): ${A_{{\rm{RRH\_}}E,\max }}$ & 8 dBi \\
maximum directional gain (MR): ${A_{{\rm{MR\_}}E,\max }}$ & 5 dBi \\
\bottomrule
\end{tabular}
\label{table3}
\end{center}
\end{table}

In order to evaluate the performance of the proposed DQN algorithm to solve P1, the following four algorithms are used as comparison schemes:
\begin{enumerate}
%$1) PNOU \ (Priority \ based \ on \ the \ number \ of \ users)
\item FBA (Fixed Beam Angle):
Consistent with the setting of the beam on the RRH in this paper, in this scheme, $\theta _B^r$ and $\phi _B^r$ are also set as fixed values, which is also the benchmark scheme used when defining the reward function. However, the random determination of these fixed values is likely to cause the RSP of MR to frequently fluctuate or stay at a low level for a long time, which makes this scheme lose its comparative significance. In the FBA scheme of this paper, the $\theta _B^r$ and $\phi _B^r$ are fixed to $0^{\circ}$. At this time, the researched train-ground communication system has higher-level RSP in most positions~\cite{He0}.

%$2) PD \ (Priority \ based \ on \ distance):
\item  $\gamma $-greedy (Greedy algorithm for adjustment of beam direction with ${\gamma _1}$ and ${\gamma _2}$ as parameters):
Considering the correlation of the optimal beam direction on adjacent location bins, the greedy algorithm shown in Algorithm~\ref{alg1} is proposed. The algorithm first initializes the environmental parameters of the researched mm-wave train-ground communication system, and set the $\theta _B^r({l_0})$ and $\phi _B^r({l_0})$ when MR is at the beginning of the railway to be $0^{\circ}$. Line 1 to Line 5 mainly explain that when MR enters the new location bin ${l_k}$, the $\theta _B^r({l_{k - 1}})$ and $\phi _B^r({l_{k - 1}})$ of the previous location bin ${l_{k - 1}}$ will be the basic values, ${\gamma _1}$ and ${\gamma _2}$ are used as the adjustment steps of the basic values, respectively. Then we can obtain the RSP of MR corresponding to all possible beam directions in this location bin ${l_k}$, and take the beam direction corresponding to the maximum RSP as the optimum. Finally, return the sequence of optimal beam directions and maximum RSP values.
\begin{algorithm}[!t]
\caption{Greedy Algorithm for Adjustment of Beam Direction with ${\gamma _1}$ and ${\gamma _2}$ as Parameters}
\label{alg1}
\hspace*{0.02in} {\bf Initialization:}
$\theta _B^r({l_0}) = 0^{\circ}$, $\phi _B^r({l_0}) = 0^{\circ}$, ${P_{r,\max}}(l)$= -200, $l$=${l_1}$,${l_2}$,...,${l_N}$.
\begin{algorithmic}[1]
\FOR{$n = 1:N$}
\FOR {${\gamma _1} =  - {\sigma _B}:{\sigma _B}: + {\sigma _B}$}
\FOR {${\gamma _2} =  - {\sigma _B}:{\sigma _B}: + {\sigma _B}$}
\STATE $\theta _B^r({l_n}) = \theta _B^r({l_{n - 1}}) + {\gamma _1}$;
\STATE $\phi _B^r({l_n}) = \phi _B^r({l_{n - 1}}) + {\gamma _2}$;
\STATE Obtain ${P_r}({l_n})$ based on $\theta _B^r({l_n})$ and $\phi _B^r({l_n})$;
\IF{${P_r}({l_n}) > {P_{r,\max }}({l_n})$}
\STATE ${P_{r,\max }}({l_n}) = {P_r}({l_n})$;
\ENDIF
\ENDFOR
\ENDFOR
\ENDFOR
\end{algorithmic}
\hspace*{0.02in} {\bf Output:} $\theta _B^r$, $\phi_B^r$ and ${P_{r,\max }}$.
%\STATE Initialize environment parameters;
%\STATE Initialize beam direction: $\theta _B^r({l_0}) = 0$, $\phi _B^r({l_0}) = 0$, ${P_{r,\max }}(l) =  - 200,l = {l_1},{l_2},...,{l_N}$;
%\STATE For $n = 1:N$;
%\STATE For ${\gamma _1} =  - {\sigma _B}:{\sigma _B}: + {\sigma _B}$
%\STATE For ${\gamma _2} =  - {\sigma _B}:{\sigma _B}: + {\sigma _B}$
%\STATE $\theta _B^r({l_n}) = \theta _B^r({l_{n - 1}}) + {\gamma _1}$;
%\STATE $\phi _B^r({l_n}) = \phi _B^r({l_{n - 1}}) + {\gamma _2}$.
%\STATE Obtain the RX power ${P_r}({l_n})$ on the MR side at the position based on $\theta _B^r({l_n})$ and $\phi _B^r({l_n})$;
%\STATE If ${P_r}({l_n}) > {P_{r,\max }}({l_n})$
%\STATE ${P_{r,\max }}({l_n}) = {P_r}({l_n})$.
%\STATE End If
%\STATE End For
%\STATE End For
%\STATE Output ${P_{r,\max }}$.
%\end{algorithmic}
\end{algorithm}

\item Q-learning:
Q-learning is one of the most commonly used algorithms in the field of RL. Compared with the proposed DQN algorithm, it has one less function approximation process, i.e., the stage of training Q-network. The Q-learning algorithm stores the reward for each state-action pairs in a Q-table. Therefore, when the state set or action set is large, it often faces the problem of high spatial complexity.

%$3) IP \ (Interior \ point \ algorithm):
\item 16-beams DQN (DQN based on codebook of 16 beams):
In order to fully observe the performance of the proposed DQN algorithm to solve P1, we do not consider the codebook. However, codebook is also an important element affecting communication performance in practice, so we use the DQN algorithm considering the codebook as a comparison scheme. According to~\cite{r12}, we consider setting the adjustment range of the $\theta _B^r$ to $[ - 1,0]^{\circ}$, taking $[ - 11,0]^{\circ}$ as the counterpart of the $\phi _B^r$. Then we can divide these two intervals into three equal parts, and thus $4 \times 4$ beam directions is formed as the codebook. Other details are consistent with the proposed DQN algorithm.
\end{enumerate}

%---------------------------
%\subsection{Simulation Results}\label{S5-2}
\subsection{Simulation Results and Discussions}\label{S5-2}
%---------------------------

%\subsubsection{Simulation System scenario}\label{S5-2-1}
%In order to have an intuitive understanding of the simulation scenario, we depict the locations of BS and MRs, as well as the locations of 200 users that randomly appear in this era when the program is executed once, as shown in Fig. \ref{fig4-3}. Among them, the red solid square above the regional center represents the mm-wave track-side BS, and the ``+" near it represents the users associated with the BS. The blue solid triangles lined up with ``one" under the center of the area represent 9 mm-wave FD MRs, and the dots nearby represent the users associated with them, but it should be noted that these users are respectively accessed through the nearest MR.

%---------------------------
\subsubsection{RSP under Different Algorithms}\label{S5-2-1}
%---------------------------

\begin{figure}[t]
\begin{center}
\includegraphics*[width=1\columnwidth,height=2.5in]{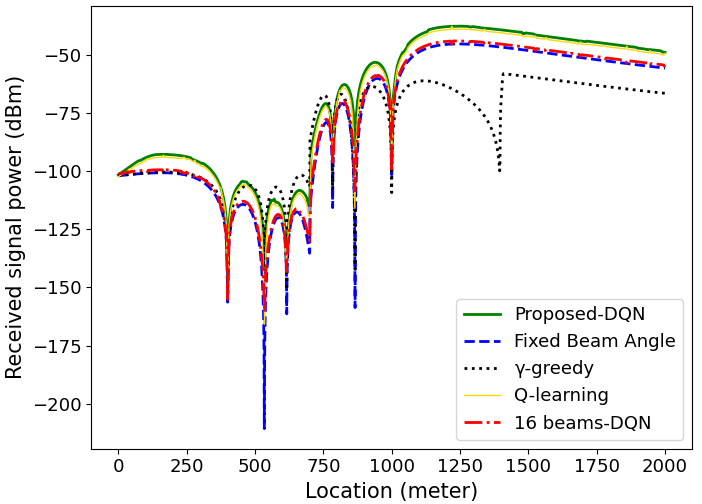}
\end{center}
\caption{RSP with MR beamforming under different algorithms.} \label{fig9}
\end{figure}

In this simulation, we use the proposed DQN algorithm and four comparison schemes to evaluate the RSP at different positions on the railway. The simulation results are shown in Fig.~\ref{fig9}.

\begin{figure}[t]
\begin{center}
\includegraphics*[width=1\columnwidth,height=2.5in]{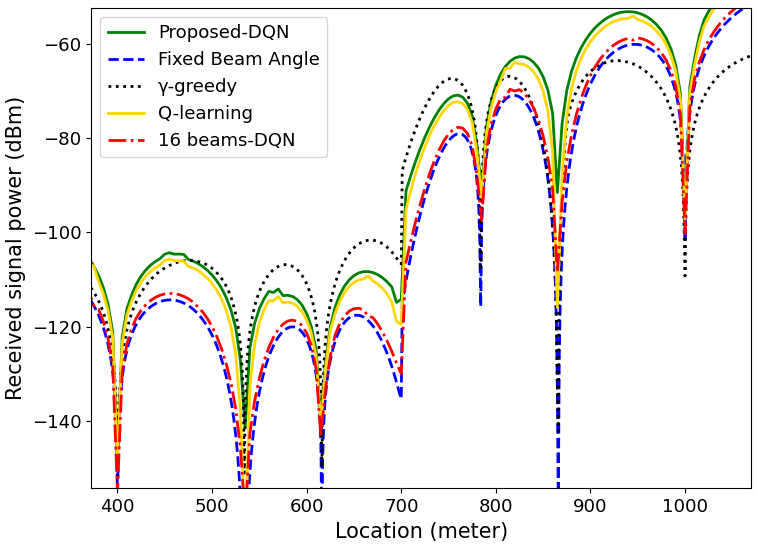}
\end{center}
\caption{RSP near the RRH with MR beamforming under different algorithms.} \label{fig9-5}
\end{figure}

It can be found that as the MR moves forward, the trends of the RSP obtained under different algorithms are similar. From overall view, the performance of the proposed DQN algorithm is the best, followed by the Q-learning algorithm, and this is in line with the previous theoretical analysis. The main problem of Q-learning is high spatial complexity, so if only from the performance, it is very close to the counterpart of the proposed algorithm.
%The codebook is not introduced into the proposed DQN algorithm, which is an ideal situation, and so it can search the optimal beam from the entire space and release the performance to the greatest extent. The 16-beams DQN limits the search range of beam to 16 optional directions, which inevitably limits the actual performance.

On the other hand, FBA and $\gamma$-greedy algorithms have poor performance compared to the other two algorithms. When MR is on the left side of RRH and is 400 to 700m away from it, the performance of the two algorithms is the same. When MR approaches the RRH from the left side and the distance is within 300m, the performance of the $\gamma$-greedy algorithm is significantly better than the FBA algorithm, and even exceeds the proposed DQN algorithm in some specific positions as shown in Fig.~\ref{fig9-5}. When MR is located near the right side of RRH, the $\gamma$-greedy algorithm still has the performance comparable to the proposed DQN algorithm, but after MR reaches about 200m to the right of RRH, the performance of the $\gamma $-greedy algorithm has experienced a significant decline, and the corresponding RSP is always lower than the FBA scheme.

It should be noted here that when using the $\gamma $-greedy algorithm, the length of the location bin $2{\sigma _D} = 1{\rm{ m}}$. The reason we set it is to consider that compared to the process of frequent interaction between the agent and the communication environment in the proposed DQN algorithm, the complexity of the greedy algorithm is lower, and so it is reasonable to adjust the beam direction more frequently. Obviously, when ${\sigma _D}$ is smaller, beam adjustment is more frequent, and the algorithm in the same setting should have better performance. However, according to Fig.~\ref{fig9}, it is clear that $\gamma $-greedy algorithm also have very poor performance on the right of the RRH. We think this is caused by the adjustment of the beam direction not keeping up with the change of the MR position, i.e., ${\gamma _1}$ and ${\gamma _2}$ should be set to larger values. This problem can be solved by adjusting the parameters, such as trying to make ${\sigma _D}$ smaller or choosing more appropriate ${\gamma _1}$ and ${\gamma _2}$.

In this simulation, with the control of program, the average reward obtained under different algorithms are returned, i.e., the average RSP gap obtained in each location bin between the FBA scheme and other algorithms. The results show that the average reward obtained by the proposed DQN algorithm is 9.29dB, while the counterpart of the 16 beams-DQN algorithm is only 1.7dB. Compared with the FBA scheme, the RSP gain of 8 times is obtained when the proposed DQN algorithm is used, and the counterpart of the 16 beams-DQN algorithm is 1.48 times. The gap between the two algorithms is clearly shown in Fig.~\ref{fig9}. As mentioned before, this is mainly caused by the limitation of the codebook on the beam search space. Therefore, it is foreseeable that adjusting the codebook can make the performance of the 16 beams-DQN gradually approach the proposed DQN algorithm, which will also be the direction of our research in the future.

%---------------------------
\subsubsection{Data Processing after DRL}\label{S5-2-2}
%---------------------------

\begin{figure*}[t]
\begin{minipage}[t]{0.5\linewidth}
\centering
\includegraphics[width=0.9\columnwidth,height=1.7in]{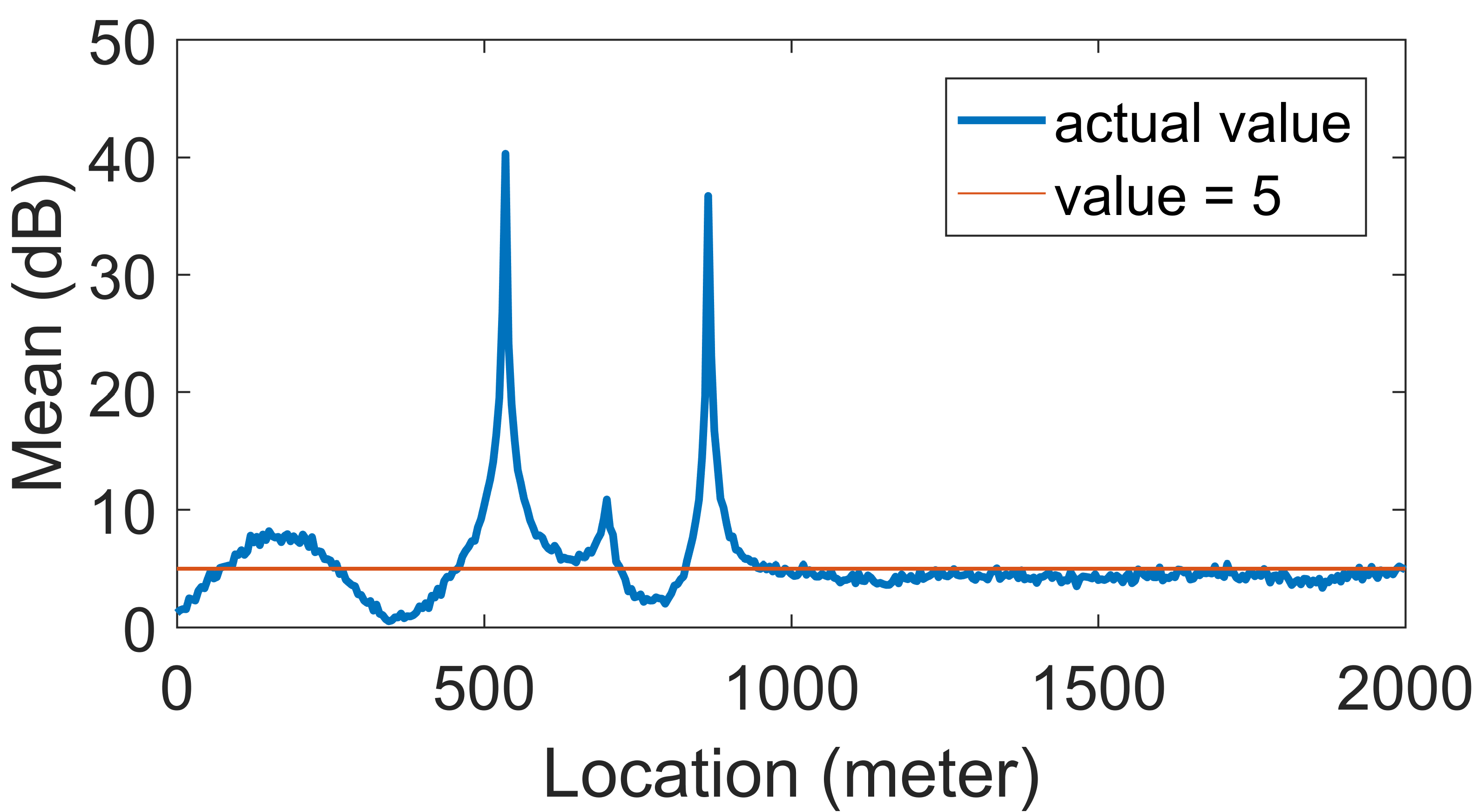}
\centerline{\small (a)}
\end{minipage}%
\begin{minipage}[t]{0.5\linewidth}
\centering
\includegraphics[width=0.9\columnwidth,height=1.7in]{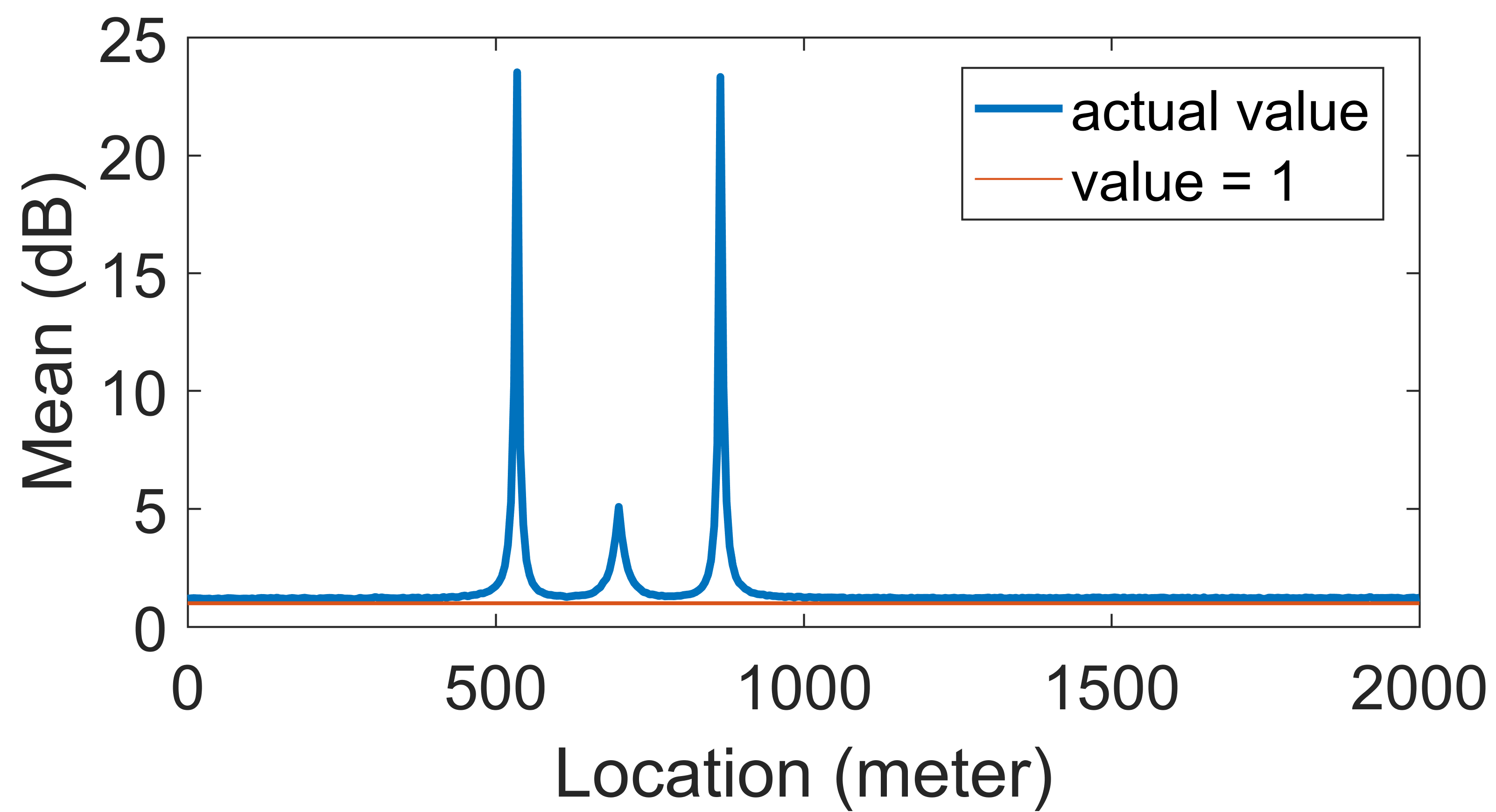}
\centerline{\small (b)}
\end{minipage}
\caption{The mean of the gap between the (a) proposed-DQN or (b) 16 beams-DQN and FBA.}
\vspace*{-3mm}
\label{fig10}
\end{figure*}

\begin{figure*}[t]
\begin{minipage}[t]{0.5\linewidth}
\centering
\includegraphics[width=0.9\columnwidth,height=1.775in]{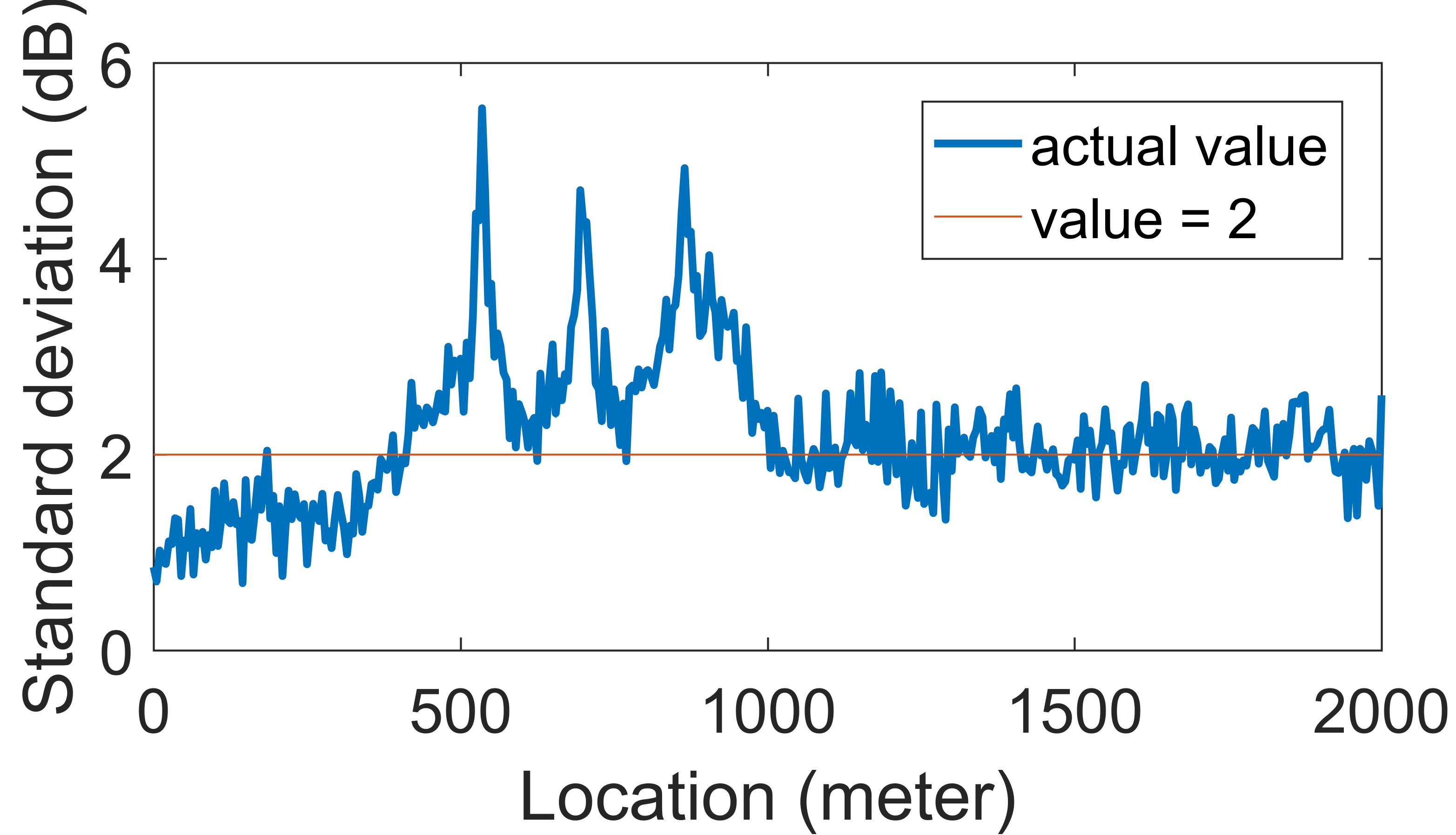}
\centerline{\small (a)}
\end{minipage}%
\begin{minipage}[t]{0.5\linewidth}
\centering
\includegraphics[width=0.9\columnwidth,height=1.7in]{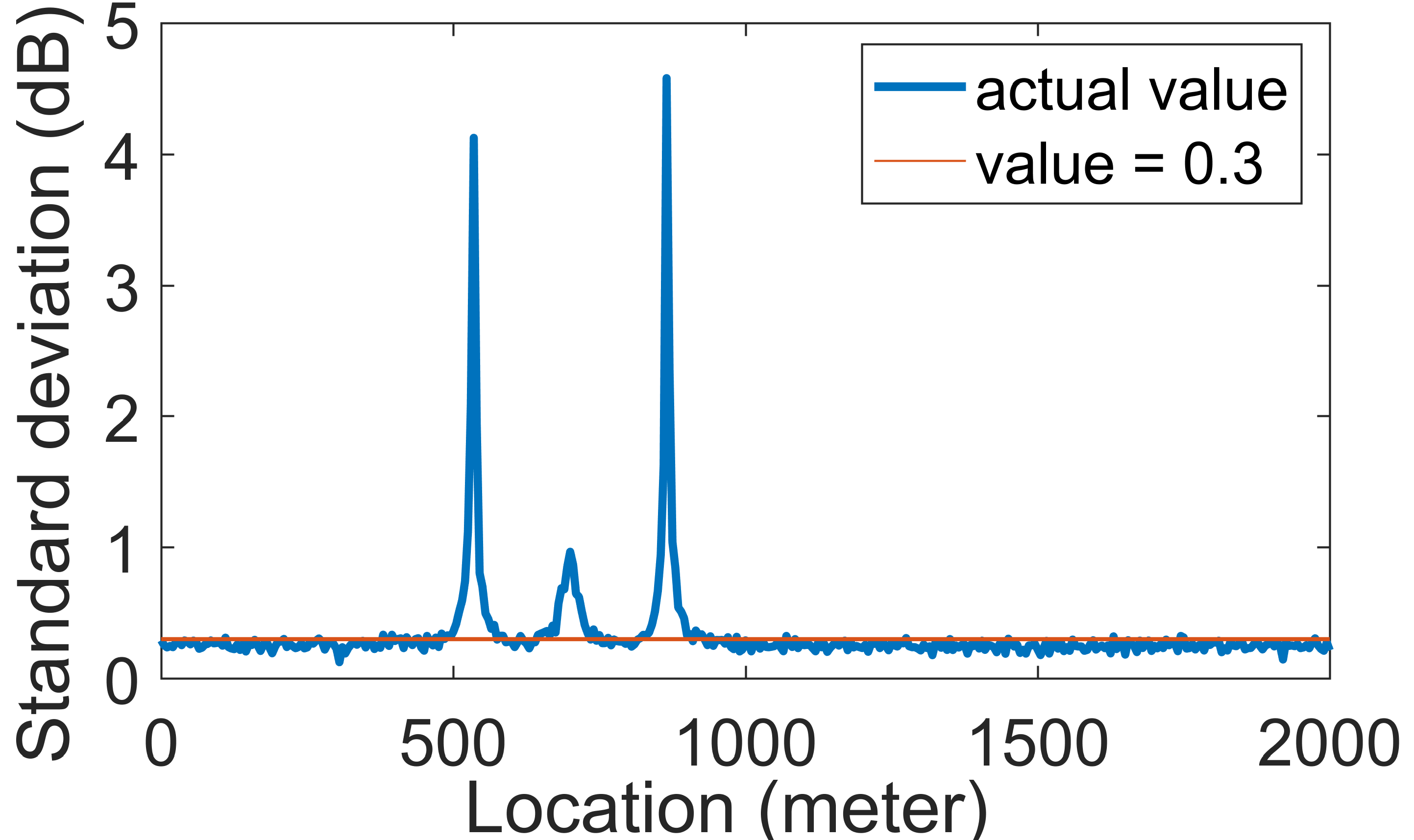}
\centerline{\small (b)}
\end{minipage}
\caption{The standard deviation of the gap between the (a) proposed-DQN or (b) 16 beams-DQN and FBA.}
\vspace*{-3mm}
\label{fig11}
\end{figure*}

\begin{figure*}[t]
\begin{minipage}[t]{0.5\linewidth}
\centering
\includegraphics[width=0.9\columnwidth,height=1.75in]{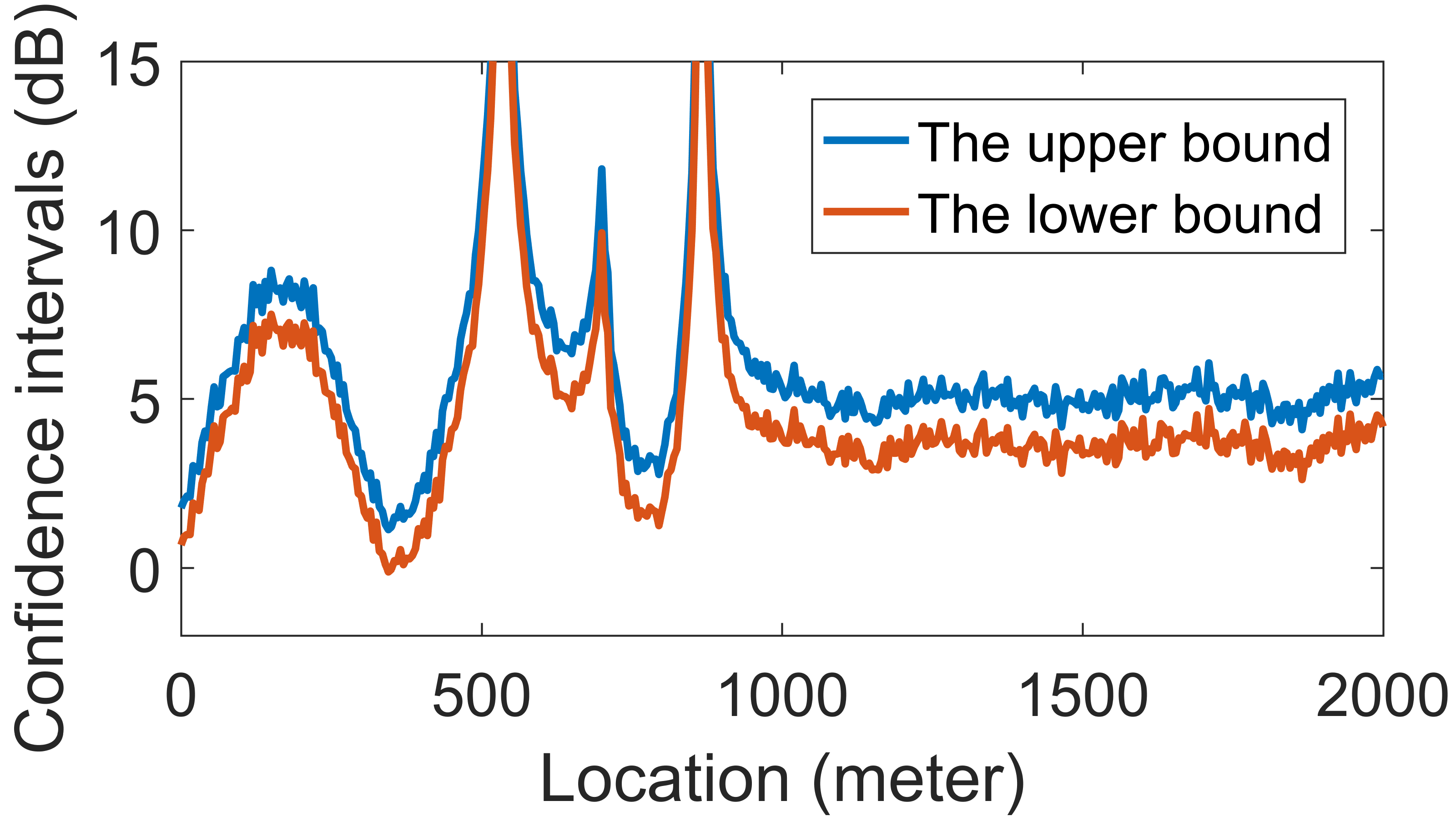}
\centerline{\small (a)}
\end{minipage}%
\begin{minipage}[t]{0.5\linewidth}
\centering
\includegraphics[width=0.9\columnwidth,height=1.7in]{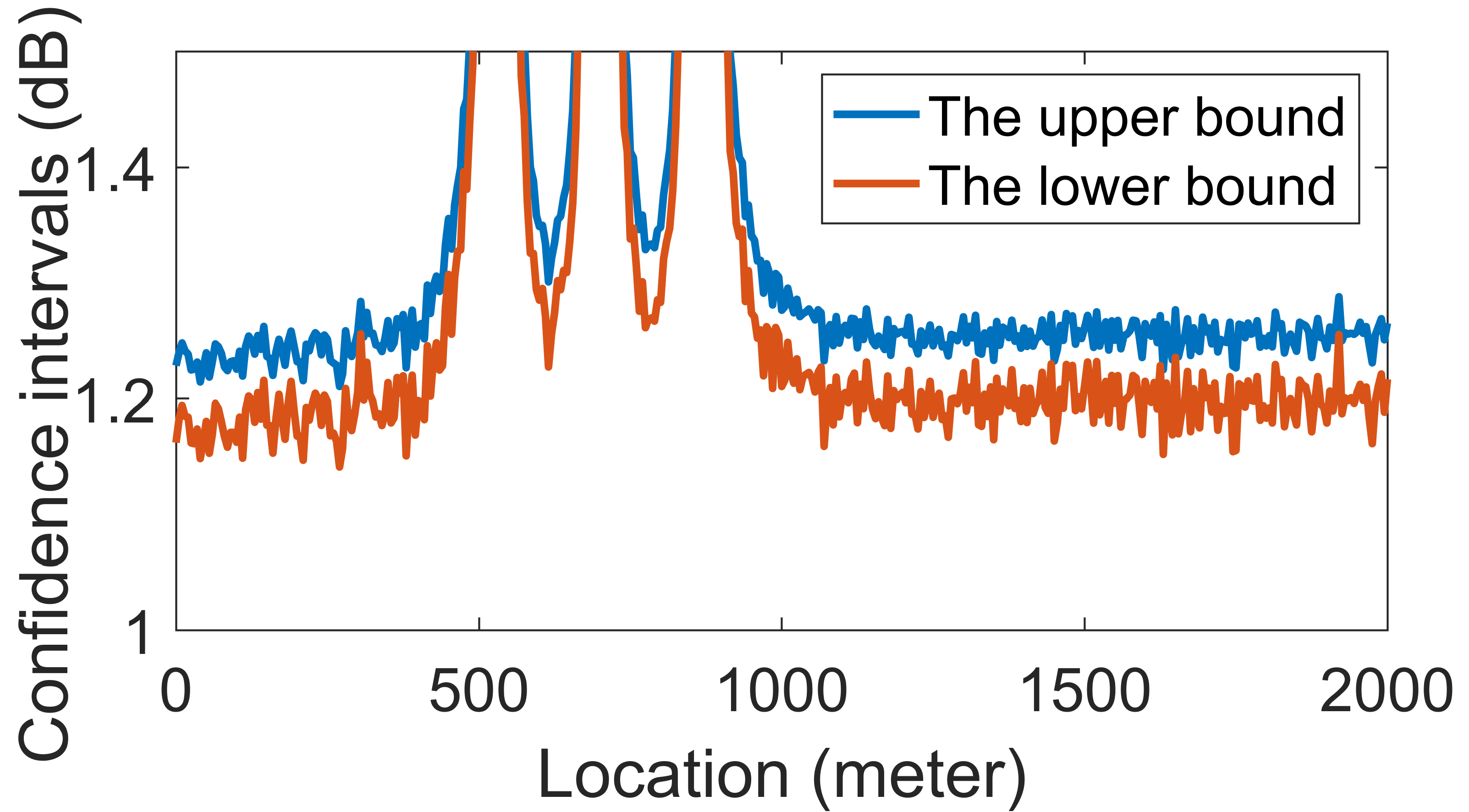}
\centerline{\small (b)}
\end{minipage}
\caption{The 95$\%$ confidence interval of the gap between the (a) proposed-DQN or (b) 16 beams-DQN and FBA.}
\vspace*{-3mm}
\label{fig12}
\end{figure*}

In order to further compare the performance of the proposed DQN algorithm and the 16 beams-DQN, after learning, the database shown in Table II is generated, the probability ${P_1}$ of ``utilization" is set to be 90$\%$, and the probability of ``exploration" is 10$\%$. We consider that the HST is continuously advancing on a 600km railway, i.e., continuously running 300 cycles on a unit of railway with a length $L$ of 2,000m, as tests after DRL. We can record the reward obtained at each location bin in these 300 tests, and count the mean, standard deviation and 95$\%$ confidence interval of these data. In addition, on each unit of railway, the RRH is located 700m from the beginning of the rail, serving the MR within the entire 2,000m. The statistical results are shown in Figs.~\ref{fig10}, \ref{fig11} and~\ref{fig12}, where Fig.~\ref{fig10} shows the mean of the gap between the two algorithms and FBA, Fig.~\ref{fig11} shows the standard deviation of gap and Fig.~\ref{fig12} shows the confidence interval of gap.

From the mean of the data, the average RSP gain at most positions under the proposed DQN algorithm is 4 to 5dB, with four peaks and three valleys, the peak is up to about 40dB, and the valley is at least a non-negative value. This proves that the proposed DQN algorithm does have a improvement of RSP for all positions, and the RSP gain obtained at most positions is 2.5 to 3.2 times that of the FBA scheme. Compared with the proposed DQN algorithm, the 16 beams-DQN constrains the beam search space, reduces the complexity and actual performance. According to Fig.~\ref{fig10}(b), the average RSP gain for most locations is between 1.3 and 2dB, which is 1.4 to 1.6 times that of the FBA scheme.

Since the codebook is not introduced into the proposed DQN algorithm, theoretically the fluctuation of data should be greater than 16 beams-DQN, which mainly occurs when the agent executes ``exploration". This is exactly the case shown in Figs.~\ref{fig11}(a) and (b). The standard deviation of the gap obtained by the proposed DQN algorithm is about 2dB. The positions before 500m correspond to smaller values, and the counterpart of position between 600m to 900m is larger, then it gradually stabilized to about 2dB. Three peaks are generated at about 550m, 700m and 850m, and the corresponding values of standard deviation are between 4dB and 6dB, indicating that the RSP near the RRH is more sensitive to the adjustment of the beam direction than the position far away from the RRH. On the other hand, the standard deviation obtained by the 16 beams-DQN is smaller, and the overall value is around 0.3dB, but there are still three peaks, corresponding to the three peaks in Fig.~\ref{fig10} and Fig.~\ref{fig11}(a). These give us some inspiration for future work. For example, if the ${\sigma _B}$ is also used as a variable parameter, a smaller ${\sigma _B}$ should be set at the positions corresponding to the peaks in Fig.~\ref{fig10} and Fig.~\ref{fig11}.

The 95$\%$ confidence interval of the data obtained by the the proposed DQN algorithm are between 0dB and 5dB, and the length of the interval are generally 1.5dB to 3dB, which correspond to the results in Fig.~\ref{fig10}(a) and Fig.~\ref{fig11}(a), respectively. The counterpart of the 16 beams-DQN are between 1.1dB and 1.4dB, and it is narrower than that of the proposed DQN algorithm, indicating that these values of reward are more concentrated, which correspond to the results in Fig.~\ref{fig10}(b) and Fig.~\ref{fig11}(b), respectively.

%In general, Fig.~\ref{fig9} and Fig.~\ref{fig9-5} prove that the proposed DQN algorithm can indeed obtain excellent RSP, and the analysis of Table IV also shows its low time-complexity. Therefore, this scheme combines high performance and low complexity. At the same time, the results shown in Fig.~\ref{fig11} and Fig.~\ref{fig12} also prove that the proposed DRL-based RX beamforming scheme has good convergence and stability. From the overall implementation process, the proposed DQN algorithm does not rely on a large number of training samples and environmental information, which effectively improves the applicability of the model. There is no massive iteration, which also makes it very suitable for HSR scenarios, and this can provide further thinking for introducing DRL into the actual mm-wave train-ground communication system.

%---------------------------
\subsubsection{Convergence Analysis}\label{S5-2-3}
%---------------------------

Unlike some previous works that used DRL to optimize traditional cellular network communications, we are not assuming that a Q-network is retrained at each location bin, and when the HST reaches a new location bin, it traverses all possible state-action pairs at this position until the algorithm converges to obtain the optimal beam direction. For the convenience of explanation, we mark this idea as ${{\rm{I}}_{\rm{1}}}$. We believe that the scheme designed based on the ${{\rm{I}}_{\rm{1}}}$ cuts off the hidden relationship in the investigated mm-wave train-ground communication system when the HST are located in adjacent location bins.

In this paper, in order to explore the hidden patterns of the mm-wave train-ground communication system as much as possible, we consider the process of the HST passing through the cell as a complete episode, and only one Q-network is trained to optimize its associated beam management process, we mark this idea as ${{\rm{I}}_2}$. In other words, when the HST is located at a fixed position, what we hope to obtain is the trend of the optimal beam angle change compared to the previous position, rather than a definite value of optimal beam angle, because the latter depends on many environmental parameters, such as the distance between the BS and rail, the height difference between the BS and MR, the average building height, etc., these undoubtedly reduce the applicability of the scheme.

After a finite number of episodes, the algorithm designed based on ${{\rm{I}}_1}$ will eventually converge and obtain the corresponding optimal beam direction of each location bin. We can easily understand this result, because in the case of a limited number of location bins $N$ and length of rail $L$, ${{\rm{I}}_1}$ is similar to the exhaustive scheme. In this paper, we designed the RX beamforming scheme based on ${{\rm{I}}_2}$. Since the reward function is related to the RSP, if the action executed by the agent is different from last episode when HST is in a forward location bin, it will affect the value of RSP at the subsequent location bin, resulting in a certain degree of fluctuation of the RSP curve. Therefore, it seems impossible to generate a completely stable strategy within a limited number of episodes, which is a natural problem of the proposed algorithm, but this does not affect the decision-making process of the agent when the HST is at the subsequent position, because the system state has not changed at this time. On the other hand, according to the results shown in Fig.~\ref{fig11} and Fig.~\ref{fig12}, the fluctuation of RSP is only about 0.3dB in most positions, that is to say, the problem mentioned above is slight after the proposed algorithm has experienced a limited number of episodes, and these results are gratifying. To a certain extent, it also shows that the proposed algorithm has a certain ability to correct the average RSP when the HST passes through the cell, and the flexibility it brings to the system is unmatched by ${{\rm{I}}_1}$.

%---------------------------
\subsubsection{Computational Complexity}\label{S5-2-4}
%---------------------------

\begin{table}[t]
\begin{center}
\caption{Computational Complexity Comparisons}
\setlength{\tabcolsep}{7mm}
\renewcommand{\arraystretch}{1.35}
\begin{tabular}{l|l}
\toprule
Name & Computational complexity \\
\midrule
Proposed-DQN & ${\rm O}(E(\frac{{{D_1}}}{B} + N)T + \bm{NT})$  \\
FBA & ${\rm O}(NT)$ \\
$\gamma $-greedy & ${\rm O}(\sigma _B^2NT)$ \\
Q-learning  & ${\rm O}((1 + E)NT)$ \\
16-beams DQN & ${\rm O}(E(\frac{{{D_2}}}{B} + N)T + \bm{NT})$ \\
\bottomrule
\end{tabular}
\label{table4}
\end{center}
\end{table}

In order to better illustrate the advantages of the proposed algorithm, we compare the computational complexity of it and the four baseline algorithms. Relevant results are shown in Table~\ref{table4}.

In the Section~\ref{S4-2}, we have analyzed in detail the computational complexity of proposed DQN algorithm, which is mainly composed of the offline and the online phase, corresponding to ${\rm{O}}(E(\frac{{{D_1}}}{B} + N)T)$ and ${\rm{O}}(NT)$.

The FBA algorithm requires MR to receive with a fixed beam, so it is the most basic scheme and has the lowest computational complexity. But according to the results shown in Fig.~\ref{fig9}, the system performance obtained by it is the worst. The $\gamma $-greedy algorithm uses the correlation of the optimal beam direction when the HST is at the front and rear positions, and determines the optimal beam direction at current position by traversing $\sigma _B^2$ directions near the optimal beam direction of the previous position at each position. Therefore, compared with FBA scheme, it has a higher complexity. According to the results shown in Fig.~\ref{fig9-5}, a higher RSP is obtained by $\sigma _B^2$-greedy algorithm near the RRH, but the RSP level is lower at the position far from the RRH, which also reflects that the performance of $\gamma $-greedy algorithm is subject to the setting of the ${\sigma _B}$, and the process of determining ${\sigma _B}$ is a trade-off between system performance and complexity.

Compared with the Q-learning algorithm, proposed algorithm has one more DQN training stage, so the computational complexity of the former is reduced to ${\rm{O}}((1 + E)NT)$. However, the DNN is introduced to solve the RL problem including large-scale state and action sets, which often requires too much storage space. Therefore, the advantage of proposed algorithm over Q-learning is reflected in the space complexity. It is conceivable that when multiple MRs are deployed on the HST, the action of adjusting the beam angle is more complicated, the state and action space of the investigated mm-wave train-ground communication system will increase geometrically, causing problems for the storage of calculation results.

The difference between 16 beams-DQN and proposed algorithm is only the size of data set D, which represents the search space of optimal beam. But when the HST is running in the actual HSR scenarios, the proposed algorithm and 16 beams-DQN algorithm are both in the online training phase. At this time, MR can directly execute the most valuable action after observing the system state, so we only need to pay attention to the complexity of them in the online phase, i.e., ${\rm{O}}(NT)$ , and it is also the lowest computational complexity we can achieve. In other words, 16 beams-DQN only reduces the computational complexity of the offline phase by reducing the beam search space.
%Combined with the results shown in Fig.~\ref{fig9}, the proposed RX beamforming scheme does have both high performance and low complexity.

In general, Fig.~\ref{fig9} and Fig.~\ref{fig9-5} prove that the proposed DQN algorithm can indeed obtain excellent RSP, and the analysis of Table IV also shows its low time-complexity. Therefore, this scheme combines high performance and low complexity. At the same time, the results shown in Fig.~\ref{fig11} and Fig.~\ref{fig12} also prove that the proposed DRL-based RX beamforming scheme has good convergence and stability. From the overall implementation process, the proposed DQN algorithm does not rely on a large number of training samples and environmental information, which effectively improves the applicability of the model. There is no massive iteration, which also makes it very suitable for HSR scenarios, and this can provide further thinking for introducing DRL into the actual mm-wave train-ground communication system.

These results prove that it is feasible to design beamforming scheme based on DRL in HSR scenarios, and they are obtained on the simulation platform based on 3GPP reference model~\cite{r1}~\cite{r2}~\cite{r3}~\cite{r11}, which will provide guidance for the landing of FR2 HST communication in the future to a certain extent.

%---------------------------
\section{Conclusions}\label{S6}
%---------------------------

Mm-wave is regarded as a promising technology adopted by train-ground communication system to meet the high data rates demand of the future smart railway system. Aiming at solving the problem of signal quality degradation caused by the high speed of HSR, this paper focused on the RX beamforming in HSR scenarios. With the purpose of maximizing average RSP, we took powerful knowledge discovery capability of AI to optimize the RX beamforming and further proposed a DQN-based scheme with a detailed definition of state, action and reward function. The proposed scheme consists of two stages. The first is the transition generating stage, which uses the Q-learning to adjust the direction of the RX beam online. The second is the parameter updating stage. At this stage, the proposed algorithm uses the stochastic gradient descent method to update the parameters of NN to make the prediction of Q-values more accurate.

Extensive simulation results demonstrated that the proposed DQN-based RX beamforming scheme can effectively improve the RSP of MR in the researched mm-wave train-ground communication system, while ensuring the low complexity and stability. For the future work, we will deeply investigate optimization of beam management under the HSR scenarios with multi-MR and multi-RRH cooperative communications.

\begin{comment}
In future work, we will
%also
consider collecting more
%personalized information on the user side before calculation to ensure that the entire train-ground communication system is more fair.
\end{comment}

%---------------------------

\bibliographystyle{IEEEtran}

\end{document}